\documentclass[12pt]{article}
\usepackage{multirow}
\usepackage{graphicx}
\graphicspath{{./figure/}}
\usepackage{multicol}
\usepackage{here}
\usepackage{comment}
\usepackage{amsmath}
\usepackage{amssymb}
\usepackage{bm}
\usepackage{subfigure}
\usepackage{cite}
\usepackage{caption}
\usepackage{authblk}
\usepackage{amsmath}
\usepackage{indentfirst}
\usepackage{calc}
\usepackage{array}
\usepackage{makecell}

\title{Evaluating marginal likelihood approximations of dose-response relationship models 
in Bayesian benchmark dose methods for risk assessment}
\author[1]{Sota Minewaki}
\author[2]{Tomohiro Ohigashi}
\author[2]{Takashi Sozu}

\affil[1]{Department of Information and Computer Technology, Tokyo University of Science, Graduate School of Engineering}
\affil[2]{Department of Information and Computer Technology, Faculty of Engineering, Tokyo University of Science}

\affil[ ]{Corresponding author. E-mail address: 4623531@ed.tus.ac.jp (Sota Minewaki)}

\date{}

\begin{document}

\maketitle
\renewcommand{\thefootnote}{}
\footnotetext{Abbreviations: BMA, Bayesian model averaging; BMD, benchmark dose; BMR, benchmark response; EPA, Environmental Protection Agency; LOAEL, lowest observable adverse effect level; MCMC, Markov chain Monte Carlo; NIEHS, National Institute of Environmental Health Sciences; NOAEL, no-observable adverse effect level; NTP, National Toxicology Program; POD, point of departure.}

\noindent
Abstract: Benchmark dose (BMD; a dose associated with a specified change in response) is used to determine the point of departure for the acceptable daily intake of substances for humans. Multiple dose-response relationship models are considered in the BMD method. Bayesian model averaging (BMA) is commonly used, where several models are averaged based on their posterior probabilities, which are determined by calculating the marginal likelihood (ML). Several ML approximation methods are employed in standard software packages, such as BBMD, \texttt{ToxicR}, and Bayesian BMD for the BMD method, because the ML cannot be analytically calculated. Although ML values differ among approximation methods, resulting in different posterior probabilities and BMD estimates, this phenomenon is neither widely recognized nor quantitatively evaluated. In this study, we evaluated the performance of five ML approximation methods: (1) maximum likelihood estimation (MLE)-based Schwarz criterion, (2) Markov chain Monte Carlo (MCMC)-based Schwarz criterion, (3) Laplace approximation, (4) density estimation, and (5) bridge sampling through numerical examples using four real experimental datasets. Eight models and three prior distributions used in BBMD and \texttt{ToxicR} were assumed. The approximation and estimation biases of bridge sampling were the smallest regardless of the dataset or prior distributions. Both the approximation and estimation biases of MCMC-based Schwarz criterion and Laplace approximation were large for some datasets. Thus, the approximation biases of the density estimation were relatively small but were large for some datasets. In terms of the accuracy of ML approximation methods, using Bayesian BMD, in which the bridge sampling is available, is preferred.\\


\newpage

\section{Introduction}
For risk assessment of chemical toxicity, the relationship between the amount of the test substance and its effects on human health (i.e., the dose-response relationship) is evaluated. To determine the ``acceptable daily intake'' of a test substance for humans, the point of departure (POD) for the dose level is determined based on animal experimental data. The no-observable adverse effect level (NOAEL) or the lowest observable adverse effect level (LOAEL) is commonly used as the POD. However, the NOAEL and LOAEL are sensitive to the sample size of animal experiments and the dose level to be evaluated \cite{r1,r2}. The United States Environmental Protection Agency (EPA), European Food Safety Authority (EFSA), and World Health Organization (WHO) recommend using the benchmark dose (BMD) method to establish POD \cite{r1,r2,r3,r4,r5}. In the BMD method, POD is estimated using a mathematical model representing the dose-response relationship, and this approach is better than the NOAEL- or LOAEL-based methods as the POD is uniquely determined by considering all dose-level data \cite{r5,r6}.

Multiple models can be used to implement the BMD method, and a candidate model for estimating BMD is selected based on statistical criteria such as goodness-of-fit statistics. The BMD estimate and 90\% confidence interval are calculated, and the lower limit of the 90\% confidence interval (benchmark dose lower confidence limit, BMDL) is used as the POD \cite{r1}. However, owing to the uncertainty in model selection, the chosen model may not be suitable for calculating the BMDL \cite{r6,r7,r8}. Therefore, model-averaging methods have attracted attention for partially resolving this uncertainty through weighted averaging of multiple models. The Bayesian model averaging (BMA) method, where multiple models are averaged based on the posterior probability, is commonly used instead of the traditional frequentist model averaging method \cite{r8,r9,r10,r11,r12,r13,r14,r15,r16,r17,r18,r19,r20}. In the BMA method, the BMD is expected to be estimated with relatively high precision if an appropriate prior distribution is assumed.

Several software packages have been developed recently for the BMD method using BMA. BBMD is a web software developed by Shao and Shapiro \cite{r15}. \texttt{ToxicR} (the successor to BMDS developed by EPA) is an R package developed by the National Institute of Environmental Health Sciences (NIEHS), the EPA, and the National Toxicology Program (NTP) \cite{r16,r17}. Bayesian BMD is newly released web software developed by EFSA, where the software is available in a local environment using the R package \texttt{BMABMDR} \cite{r18,r19}. The type of the response variable, dose-response relationship model, default prior distribution, derivation method of the posterior distribution of the BMD, and calculation method for posterior probability used in the three software packages are summarized in Table \ref{tb:sum}.

\begin{table}[H]
    \caption{Differences among BBMD, \texttt{ToxicR}, and Bayesian BMD}
    \label{tb:sum}
    \begin{tabular}{>{\centering\arraybackslash}m{0.22\textwidth-2\tabcolsep}>{\centering\arraybackslash}m{0.24\textwidth-2\tabcolsep}>{\centering\arraybackslash}m{0.24\textwidth-2\tabcolsep}>{\centering\arraybackslash}m{0.3\textwidth-2\tabcolsep}}
    \hline
        Software & BBMD & \texttt{ToxicR} &  Bayesian BMD \\ \hline
        Developer & Kan Shao, Andy Shapiro & NIEHS, EPA, NTP & EFSA \\ \hline
        Platform & Web application & R package & Web application and R package \texttt{BMABMDR}  \\ \hline
        Type of the response variable  &  Continuous, quantal, and ordered categorical &  Continuous and quantal &  Continuous and quantal \\ \hline
        Dose-response relationship model & 8 models (Table \ref{tb:model_prior}) & 8 models (Table \ref{tb:model_prior}) and gamma model  & 8 models (see Supplementary Material) \\ \hline
        Default prior distribution & Non-informative prior (Table \ref{tb:model_prior}) & Informative prior (Table \ref{tb:model_prior})  & Normal distribution and modified PERT distribution (see Supplementary Material)  \\ \hline
        Derivation method for the posterior distribution of the BMD & MCMC  & Laplace approximation and MCMC  & Laplace approximation and MCMC\\ \hline
        Calculation method for posterior probability  & MCMC-based Schwarz criterion & Laplace approximation & Laplace approximation and bridge sampling \\ \hline
    \end{tabular}
    \caption*{BMD: Benchmark dose; EPA: Environmental Protection Agency; MCMC: Markov chain Monte Carlo; NIEHS: National Institute of Environmental Health Sciences; NTP: National Toxicology Program; PERT: Program Evaluation Review Technique}
\end{table}

The available models in BBMD and \texttt{ToxicR} differ from those in Bayesian BMD. BBMD and \texttt{ToxicR} use the original model parameters directly in their prior distributions, whereas Bayesian BMD uses prior distributions after transforming the model parameters into ``natural parameters'' that can be easily interpreted by users. However, the models and prior distributions of Bayesian BMD have not been sufficiently discussed in previous studies using the BMD method. Thus, although Bayesian BMD includes most of the options in BBMD and \texttt{ToxicR}, these software packages are used differently depending on the user's preference. Regardless of the selected software, the marginal likelihood (ML) is calculated to obtain the posterior probability of each model. However, ML estimation involves complex integration, and the analytical calculation of ML is challenging. Therefore, several methods (see Section 2) have been proposed for approximating the ML (hereafter referred to as ML approximation methods). Although the ML values differ among approximation methods, resulting in different posterior probabilities and BMD estimates, the methods are not  quantitatively evaluated.

We aimed to evaluate the performance of different ML approximation methods and their effects on BMD estimation using numerical examples. In the numerical examples, we considered the models and prior distributions employed in BBMD and \texttt{ToxicR} because the interpretation of the obtained results would be complex and difficult if the conditions of Bayesian BMD were included.

\section{Method}
\subsection{Bayesian BMD method}
In the Bayesian BMD method, the posterior distributions of all models are calculated after parameter estimation. The posterior distribution of BMD is then obtained using the BMA method. In this study, only dichotomous datasets were used.

\subsubsection{Estimation of model parameters}
A dichotomous dataset $D$ containing the following data was considered: dose level $d_i$, number of animals $n_i$, and number of animals showing the adverse response $y_i$ for each $i$th dose group. The number of animals showing the adverse response for each dose group was assumed to follow a binomial distribution $y_i \sim \text{Bin}(n_i, p_i)$, where $p_i = f(d_i \mid \theta)$ represents the model-predicted response rate, $f(\cdot)$ is the model, and $\theta$ is the model parameters. To estimate the parameters, the likelihood function $P(D \mid \theta)$ was defined as follows:

\[
P(D \mid \theta) = \prod_i \binom{n_i}{y_i} p_i^{y_i} (1 - p_i)^{n_i - y_i}
\]
\noindent
In the Bayesian BMD method, the posterior distribution of model parameters $P(\theta \mid D)$ is obtained from the prior distribution of model parameters $P(\theta)$ and the likelihood function $P(D \mid \theta)$ using Bayes' theorem. However, posterior distributions cannot often be obtained analytically, and methods such as the Markov chain Monte Carlo (MCMC) method are used instead\cite{r15}.

\subsubsection{Estimation of BMD}
In the BMD method, the benchmark response (BMR) is defined as the target level of change in the response relative to the response at zero dosage in the model, and the dose corresponding to the BMR is the BMD. For dichotomous datasets, the following definition, called ``extra risk,'' is recommended to reflect such a change \cite{r1,r2,r3,r4}.

\begin{equation}
    \label{eq:1}
    \text{BMR} = \frac{f(\text{BMD}) - f(0)}{1 - f(0)}
\end{equation}

\noindent
In the Bayesian BMD method, the posterior sampling of the BMD is obtained by applying posterior sampling of the parameters to Equation \eqref{eq:1}. The median of the posterior sampling of BMD was used as the BMD estimate, and the lower 5th percentile was the BMDL \cite{r15}. In this study, the BMR was set to 0.1 according to the EFSA's guidance \cite{r4}.

\subsubsection{BMA method}
In the BMA method, the posterior distributions of BMD for each model are weighted and averaged based on the posterior probability of each model \cite{r21, r22}. Assuming that the $k$th model is $M_k$ ($k = 1, 2, ..., K$), the posterior distribution of the BMD after applying the BMA method, $P(\text{BMD} \mid D)$, is:

\begin{equation}
    \label{eq:2}   
    P(\text{BMD} \mid D) = \sum_{k=1}^{K} P(\text{BMD} \mid M_k, D) P(M_k \mid D),
\end{equation}
\noindent
where $P(\text{BMD} \mid M_k, D)$ is the posterior distribution of the BMD of the $k$th model, and $P(M_k \mid D)$ is the posterior probability of the $k$th model. Here, $P(M_k \mid D)$ in Equation \eqref{eq:2} is:

\begin{equation}
    \label{eq:3}
    P(M_k \mid D) = \frac{P(D \mid M_k) P(M_k)}{\sum_{t=1}^{K} P(D \mid M_t) P(M_t)},
\end{equation}

\noindent
where $P(M_k)$ is the prior probability of the $k$th model. If $K$ models exist, as is common, then $P(M_k) = \frac{1}{K}$ \cite{r13, r22}. $P(D \mid M_k)$ in Equation \eqref{eq:3} is given by:

\begin{equation}
    \label{eq:4}
    P(D \mid M_k) = \int P(D \mid \theta_k, M_k) P(\theta_k \mid M_k) d\theta_k,
\end{equation}

\noindent
where $\theta_k$ are the $k$th model parameters, $P(\theta_k \mid M_k)$ is the prior distribution of the $k$th model parameters, and $P(D \mid \theta_k, M_k)$ is the likelihood. Because the integral for Equation \eqref{eq:4} cannot be calculated analytically in many cases, an approximation method is often used.

\subsection{Models, prior distributions of parameters, and ML approximation method}
\subsubsection{Models and prior distributions}
Both the EPA and EFSA recommend using a biologically based model that describes the toxicokinetics and toxicodynamics of chemicals as the ideal approach. However, such models are rare \cite{r1,r2}. In practice, multiple models available in biostatistics and toxicology are used \cite{r6}. In this study, we used eight dose-response relationship models that are available in both BBMD and \texttt{ToxicR} (Table \ref{tb:model_prior}) \cite{r15,r16,r17}.

\begin{table}[H]
    \caption{Dose-response relationship models used in this study and default prior distributions of BBMD and \texttt{ToxicR}. Here, $d$ is the dose and $\Phi(\cdot)$ is the cumulative standard normal distribution.}
    \label{tb:model_prior}

    \begin{tabular}{>{\centering\arraybackslash}m{0.34\textwidth-2\tabcolsep}>{\centering\arraybackslash}m{0.17\textwidth-2\tabcolsep}>{\centering\arraybackslash}m{0.235\textwidth-2\tabcolsep}>{\centering\arraybackslash}m{0.255\textwidth-2\tabcolsep}}
    \hline
        Models & Constraints & BBMD non-informative $\text{prior}^{*}$ &  \texttt{ToxicR} informative $\text{prior}^{*}$ \\ \hline
        Logistic : $\frac{1}{1 + \exp{(-\alpha - \beta d)}}$& $0 < \beta$ & $\alpha \sim \text{Unif}(-50,50)$ $\beta \sim \text{Unif}(0,100)$ & $\alpha \sim \text{N}(0,1)$ $\beta \sim \text{LN}(0,2)$ \\ \hline
        Probit : $\Phi(\alpha + \beta d)$ & $0 < \beta$ & $\alpha \sim \text{Unif}(-50,50)$ $\beta \sim \text{Unif}(0,100)$ & $\alpha \sim \text{N}(0,1)$ $\beta \sim \text{LN}(0,2)$ \\ \hline
        Q-linear : $\beta + (1-\beta)(1-\exp{(-\alpha d)})$ &  $0 < \alpha$ $0 \leq \beta \leq 1$ &  $\alpha \sim \text{Unif}(0,100)$ $\beta \sim \text{Unif}(0,1)$ & $\alpha \sim \text{LN}(0,1)$ $\text{logit}(\beta) \sim \text{N}(0,2)$ \\ \hline
        Weibull : $\gamma + (1-\gamma)$\newline$(1-\exp{(-\beta\exp{(\alpha)})})$ & $0 < \alpha, \beta$  $0 \leq \gamma \leq 1$ &  $\alpha \sim \text{Unif}(0,50)$ $\beta \sim \text{Unif}(0,15)$ $\gamma \sim \text{Unif}(0,1)$ & $\alpha \sim \text{LN}(0.4243,0.5)$ $\beta \sim \text{LN}(0,1.5)$  $\text{logit}(\gamma) \sim \text{N}(0,2)$ \\ \hline
        Multi-stage2 : $\gamma + (1-\gamma)$\newline$(1-\exp{(-\alpha d -\beta d^2)})$ & $0 < \alpha, \beta$  $0 \leq \gamma \leq 1$ & $\alpha \sim \text{Unif}(0,100)$ $\beta \sim \text{Unif}(0,100)$ $\gamma \sim \text{Unif}(0,1)$ & $\alpha \sim \text{LN}(0,0.5)$ $\beta \sim \text{LN}(0,1)$  $\text{logit}(\gamma) \sim \text{N}(0,2)$ \\ \hline
        Log Logistic : $\gamma + \frac{1-\gamma}{1+\exp{(-\alpha - \beta \log{d}})}$ & $0 < \beta$  $0 \leq \gamma \leq 1$ & $\alpha \sim \text{Unif}(-5,15)$ $\beta \sim \text{Unif}(0,15)$ $\gamma \sim \text{Unif}(0,1)$ & $\alpha \sim \text{N}(0,1)$ $\beta \sim \text{LN}(0.6931,0.5)$ $\text{logit}(\gamma) \sim \text{N}(0,2)$ \\ \hline
        Log Probit : $\gamma + (1-\gamma)\Phi(\alpha + \beta \log{d})$ & $0 < \beta$  $0 \leq \gamma \leq 1$ & $\alpha \sim \text{Unif}(-5,15)$ $\beta \sim \text{Unif}(0,15)$ $\gamma \sim \text{Unif}(0,1)$ & $\alpha \sim \text{N}(0,1)$ $\beta \sim \text{LN}(0.6931,0.5)$  $\text{logit}(\gamma) \sim \text{N}(0,2)$ \\ \hline
        Dichotomous-Hill : $\delta\left( \gamma + \frac{1-\gamma}{1+\exp{(-\alpha - \beta \log{d}})} \right)$ &  $0 < \beta$  $0 \leq \gamma, \delta \leq 1$ & $\alpha \sim \text{Unif}(-5,15)$ $\beta \sim \text{Unif}(0,15)$ $\gamma \sim \text{Unif}(0,1)$  $\delta \sim \text{Unif}(-3,3.3)$ & $\alpha \sim \text{N}(0,1)$ $\beta \sim \text{LN}(0.6931,0.5)$  $\text{logit}(\gamma) \sim \text{N}(-1,2)$  $\text{logit}(\delta) \sim \text{N}(0,3)$ \\ \hline
    \end{tabular}
    \caption*{*BBMD non-informative prior: non-informative prior using a uniform distribution, the default prior for BBMD; \texttt{ToxicR} informative prior: an informative prior, the default for \texttt{ToxicR}}
\end{table}

Several studies have been conducted on the constraints for power parameters ($\alpha$ in Weibull, log logistic, and log probit and $\beta$ in dichotomous-Hill) \cite{r1, r2, r5, r6}. However, these constraints have been reported to have little effect on the estimation of power or other parameters even when a uniform distribution is used for the prior distribution \cite{r23}. Therefore, the constraints listed in Table \ref{tb:model_prior} were used.

The prior probability of each model was set to $P(M_k) = \frac{1}{8}$ for the eight models, as in previous research \cite{r13, r16}. Three prior distributions were set for the parameters of each model: a non-informative prior using a uniform distribution (the default prior for BBMD, hereafter referred to as the BBMD non-informative prior) \cite{r15}; an informative prior (the default for \texttt{ToxicR}, hereafter referred to as the \texttt{ToxicR} informative prior) \cite{r16, r17}; and an informative prior based on previous data \cite{r15, r23, r24}. The results and setup for the informative prior based on previous data are described in Section S1 of the Supplementary materials.

In this study, we considered the models and prior distributions employed in BBMD and \texttt{ToxicR} because the interpretation of the obtained results would be complex and difficult if the conditions of Bayesian BMD were included. Models and prior distributions employed in Bayesian BMD are described in Section S2 of the Supplementary materials.

\subsubsection{ML approximation method}
We compared five ML approximation methods used in previous studies \cite{r8,r9,r10,r11,r12,r13,r14,r15,r16,r17,r18,r19,r20} on BMD method using BMA.\\

\noindent
(1) Schwarz criterion\\
\indent
The Schwarz criterion approximates the log-transformed ML, also known as the Bayesian Information Criterion (BIC) \cite{r25}, and is calculated as follows:

\begin{equation}
    \label{eq:5}
    \log(P(D \mid M_k)) \approx \hat{l}_k - \frac{q_k}{2} \log(n) = -\frac{1}{2} \text{BIC},
\end{equation}

\noindent
where $\hat{l}_k$ is the maximum log-likelihood, $q_k$ is the number of parameters, and $n$ is the number of dose-level groups. Two Schwarz criterion are used for BMA in the BMD method. With the first, the maximum log-likelihood (in Equation \eqref{eq:5}) is estimated using the maximum likelihood method \cite{r8, r13} without using prior or posterior distributions (hereafter referred to as the MLE-based Schwarz criterion). With the second, the log-likelihood and Equation \eqref{eq:5} are calculated for each posterior sampling, and the results are substituted into Equation \eqref{eq:3} to obtain the posterior probability of each model for each sample. The averages of the posterior probabilities for each posterior sampling for each model are then used for the BMA (hereafter referred to as the MCMC-based Schwarz criterion) \cite{r15}. In BBMD, the MCMC-based Schwarz criterion is implemented.\\

\noindent
(2) Laplace approximation\\
\indent
The Laplace approximation is implemented in \texttt{ToxicR} and Bayesian BMD \cite{r16,r17,r18, r19, r26}. In general, the Laplace approximation performs faster than other marginal likelihood approximation methods. However, the Laplace approximation fails in certain settings, and it is not known when these failures occur \cite{r19}. The method is calculated as follows:

\[
P(D \mid M_k) \approx (2\pi)^{\frac{q_k}{2}}  \begin{vmatrix} \hat{\Sigma} \end{vmatrix}   ^{\frac{1}{2}} P(D \mid \hat{\theta}_k, M_k) P(\hat{\theta}_k \mid M_k),
\]

\noindent
where $\hat{\theta}_k$ is the maximum a posteriori (MAP) estimate of the posterior distribution of parameters and $\hat{\Sigma}$ is the inverse of the negative Hessian matrix of the posterior distribution, evaluated at $\hat{\theta}_k$. We computed the second-order partial derivatives, the elements of the Hessian matrix, by numerical differentiation, as in \texttt{ToxicR}. Let $f(\boldsymbol{\theta}) = P(D \mid \boldsymbol{\theta}, M_k)P(\boldsymbol{\theta} \mid M_k)$, the numerical derivative of the second-order partial derivative is:

\begin{align*}
\frac{\partial^2 f(\alpha, \beta)}{\partial \alpha^2} = \frac{1}{12 h_{\alpha}^2} \left\{ -f(\alpha + 2h_{\alpha}, \beta) + 16f(\alpha + h_{\alpha}, \beta) - 30f(\alpha, \beta) \right.\\
\left. + 16f(\alpha - h_{\alpha}, \beta) - f(\alpha - 2h_{\alpha}, \beta) \right\}
\end{align*}

\begin{align*}
\frac{\partial^2 f(\alpha, \beta)}{\partial \alpha \partial \beta} = \frac{1}{4 h_{\alpha} h_{\beta}} \left\{ f(\alpha + h_{\alpha}, \beta + h_{\beta}) - f(\alpha + h_{\alpha}, \beta \right. - h_{\beta})\\
\left. - f(\alpha - h_{\alpha}, \beta + h_{\beta}) + f(\alpha - h_{\alpha}, \beta - h_{\beta}) \right\}
\end{align*}

\noindent
where $\alpha, \beta \in \boldsymbol{\theta}$ and $h_{\alpha}, h_{\beta}$ are small values. We set $h_{\alpha}$ and $h_{\beta}$ to (absolute value of the estimated parameters) $\times 10^{-\frac{16}{3}}$, as in \texttt{ToxicR}.\\

\noindent
(3) Density estimation\\
\indent
The density estimation approximates $P(D \mid M_k)$ using Bayes' theorem:$P(\theta_k \mid D) = \frac{P(D \mid \theta_k, M_k) P(\theta_k \mid M_k)}{P(D \mid M_k)}$ \cite{r20,r21}
\noindent
and is calculated as follows:

\begin{equation}
    \label{eq:6}
    P(D \mid M_k) \approx \frac{P(D \mid \hat{\theta}_k, M_k) P(\hat{\theta}_k \mid M_k)}{\hat{P}(\hat{\theta}_k \mid D)},
\end{equation}

\noindent
where, $\hat{\theta}_k$ are the estimates of parameters, such as the mean value, and $\hat{P}(\hat{\theta}_k \mid D)$ is the posterior distribution of parameters. We estimated $\hat{P}(\hat{\theta}_k \mid D)$ after posterior sampling of the parameters using the \texttt{kde} function of the \texttt{ks} package in R, which performs multidimensional kernel density estimation, with default settings \cite{r27}.\\

\noindent
(4) Bridge sampling\\
\indent
The bridge sampling is implemented in Bayesian BMD \cite{r18, r19}. Approximation of the marginal likelihood can be computed by Monte Carlo integration, but this may be inefficient. There are improved methods like importance sampling and generalized harmonic mean \cite{r28,r29,r30,r31}. These methods sample from the proposed distribution instead of the prior distribution allowing for efficient sampling but imposing strong constraints on the tail behavior of the proposed distribution. The bridge sampling, which relaxes these constraints by linking the proposal distribution with the posterior distribution using a bridge function, has been proposed and applied to the BMD method \cite{r28,r29,r30,r31}. The bridge sampling is calculated as follows:

\begin{align*}
    P(D \mid M_k) &\approx 
    \frac{
    \frac{1}{n_2} \sum_{j=1}^{n_2} h(\tilde{\theta}_{kj}) P(D \mid \tilde{\theta}_{kj}, M_k) P(\tilde{\theta}_{kj} \mid M_k)}{\frac{1}{n_1} \sum_{i=1}^{n_1} h(\theta^*_{ki}) g(\theta^*_{ki})},  \\
    &\hspace{18pt} \quad\theta^*_{ki} \sim g(\theta), \quad\tilde{\theta}_{kj} \sim P(\theta_k \mid D, M_k),
\end{align*}

\noindent
where $g(\theta)$ is the proposal distribution and $h(\theta)$ is the bridge function. It is known that the computational efficiency improves when the proposal distribution and the posterior distribution overlap. In WARP-III, one of the methods for achieving this overlap, the proposal distribution is a multivariate standard normal distribution, and the posterior sampling is made close to the proposal distribution \cite{r32}. We approximated $P(D \mid M_k)$ after posterior sampling of the parameters using the \texttt{bridgesampling} package in R, which implements the bridge function that minimizes the relative mean squared error, as proposed by Meng and Wong in 1996, along with the bridge sampling using WARP-III \cite{r33}.

\section{Numerical example}
The computational environment was a desktop PC with an 11th Gen Intel Core i5-1135G7 CPU (2.40GHz) and 16 GB of memory. R version 4.2.0 (R Core Team, 2022) was used for statistical analysis.

\subsection{Settings}
We selected the commonly observed typical four datasets listed in Table \ref{tb:dataset} from the 518 datasets used by Shao and Shapiro (2018) \cite{r15}. Approximately half of the 518 datasets (244 datasets; 47.1\%) had zero response at zero dosage; hence, two of the four datasets were selected to have zero response at zero dosage. 

\begin{table}[H]
    \caption{Details of the four datasets}
    \label{tb:dataset}
    \begin{tabular}{>{\centering\arraybackslash}m{0.27\textwidth-2\tabcolsep}>{\centering\arraybackslash}m{0.08\textwidth-2\tabcolsep}>{\centering\arraybackslash}m{0.08\textwidth-2\tabcolsep}>{\centering\arraybackslash}m{0.08\textwidth-2\tabcolsep}>{\centering\arraybackslash}m{0.08\textwidth-2\tabcolsep}>{\centering\arraybackslash}m{0.08\textwidth-2\tabcolsep}>{\centering\arraybackslash}m{0.01\textwidth-2\tabcolsep}>{\centering\arraybackslash}m{0.07\textwidth-2\tabcolsep}>{\centering\arraybackslash}m{0.08\textwidth-2\tabcolsep}>{\centering\arraybackslash}m{0.08\textwidth-2\tabcolsep}>{\centering\arraybackslash}m{0.08\textwidth-2\tabcolsep}}
     & \multicolumn{5}{c}{Dataset 1} & & \multicolumn{4}{c}{Dataset 2} \\ \cline{2-6} \cline{8-11}
     Dose level & 0 & 0.25 & 0.50 & 1.00 & & & 0 & 0.10 & 0.334 & 1.00 \\ \hline
     Number of animals & 15 & 30 & 29 & 16 & & & 48 & 46 & 46 & 41 \\
     Number of animals which showed the adverse response (\%) & 0\par(0) & 5\par(16.7) & 26\par(89.7) & 16\par(100) & & & 0\par(0) & 1\par(2.2) & 4\par(8.7) & 15\par(36.6) \\ \hline
     & \multicolumn{5}{c}{Dataset 3} & & \multicolumn{4}{c}{Dataset 4} \\ \cline{2-6} \cline{8-11}
     Dose level & 0 & 0.167 & 0.332 & 0.665 & 1.00 & & 0 & 0.25 & 0.50 & 1.00 \\ \hline
     Number of animals & 27 & 28 & 28 & 28 & 29 & & 20 & 20 & 20 & 20 \\
     Number of animals which showed the adverse response (\%) & 9\par(33.3) & 20\par(71.4) & 24\par(85.7) & 27\par(96.4) & 29\par(100) & & 1\par(5.0) & 2\par(10.0) & 5\par(25.0) & 11\par(55.0) \\ \hline
    \end{tabular}
\end{table}

\begin{figure}[H]
    \centering
    \includegraphics[width=0.9\textwidth]{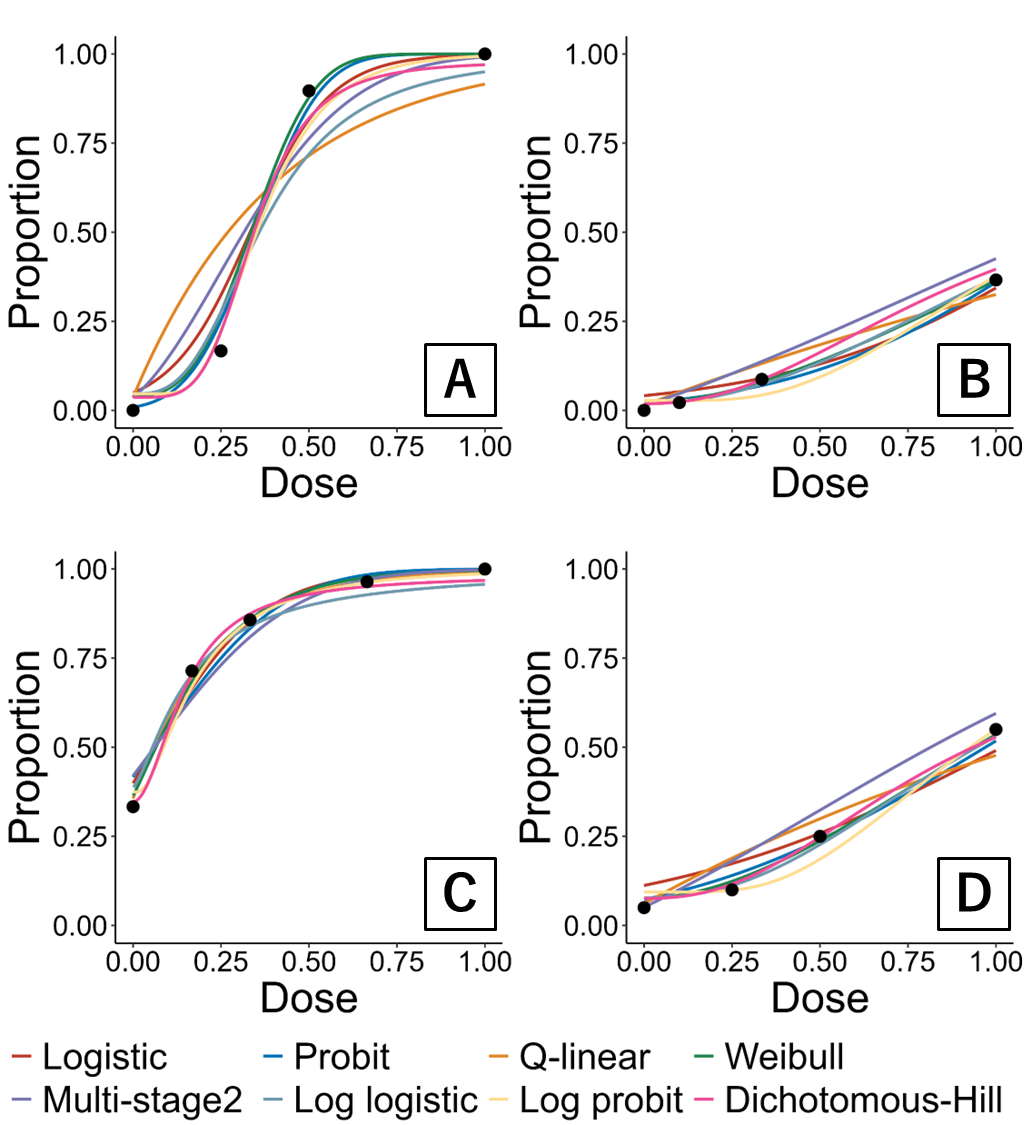}
    \caption{Scatter plots of the four datasets fitted using the eight selected models. All models were fitted using \texttt{ToxicR} informative prior. The parameter estimates were the averages estimated from the posterior sampling using \texttt{rstan}. Panel A, Dataset 1; Panel B, Dataset 2; Panel C, Dataset 3; and Panel D, Dataset 4.}
    \label{fig:1}
\end{figure}

The four datasets fitted to the eight models using the \texttt{ToxicR} informative prior are shown in Figure \ref{fig:1}, where the parameter estimates were averaged from posterior sampling using \texttt{rstan}.\\

A sigmoid dose-response relationship with zero response at zero dosage was observed for Dataset 1 (Study No. 413). A sublinear dose-response relationship with zero response at zero dosage was observed for Dataset 2 (Study No. 47), and the proportion of animals which showed the adverse response at a high dose was less than 40\%. A supralinear dose-response relationship was observed for Dataset 3 (Study No. 42), with animals showing the adverse response at zero dosage and a relatively high proportion of animals showing the response at lower doses. A linear dose-response relationship with a response at zero dosage was observed for Dataset 4 (Study No. 373), and the proportion at high doses was greater than 50\%.

\subsection{Reference values and performance evaluation}
For the four datasets, three prior distributions and five approximation methods were used to compute the ML (as in Equation \eqref{eq:3}), the BMD estimate, and the BMDL.

Following the procedure of previous study \cite{r34}, we sampled from the prior distribution of the model parameters and used the Monte Carlo integral as follows to calculate the reference value of ML in Equation \eqref{eq:3}:

$$P(D \mid M_k) \approx \frac{1}{N}\sum^N_{i=1}P(D \mid \theta_{ki}, M_k), \hspace{10pt} \theta_{ki} \sim P(\theta_k \mid M_k),$$

\noindent
where $N = 10^9$ is the number of samples of parameter $\theta_{ki}$. A smaller difference between the approximation and reference value (approximation bias) indicates better approximation performance.

Posterior samples of the parameters obtained using \texttt{rstan} \cite{r35} (three chains, 11000 iterations, and 1000 warm-ups) were used for the MCMC-based Schwarz criterion, the density estimation, and the bridge sampling.

The BMD estimate and BMDL were computed based on the BMA method (Equation \eqref{eq:2}) using three prior distributions and five approximation methods. The posterior probability $P(M_k \mid D)$ (Equation \eqref{eq:3}) for the model in Equation \eqref{eq:2} was calculated by approximating the MLs obtained using the methods described in Section 3.3. $P(\text{BMD} \mid M_k, D)$ was computed by posterior sampling of BMD using the posterior sample of parameters using \texttt{rstan}. The median posterior BMD sampling estimate was used as the BMD estimate. the lower 5th percentile was used as the BMDL. The BMD and BMDL reference values were computed using the ML reference value. Bias in BMD was defined as the difference between the BMD estimate and reference value.

\subsection{ML estimates and weights of the model}
Because the characteristics of the ML estimates were similar among the four datasets, only the results for Datasets 1 and 4 are shown in Tables \ref{tb:result4} and \ref{tb:result5}, respectively. The results of Datasets 1 and 4 when using the informative prior and those of Datasets 2 and 3 are included in Section S3.1 of the Supplementary materials. The MCMC-based Schwarz criterion could not be used to approximate the ML in Equation \eqref{eq:4} but could be used to approximate the posterior probability $P(M_k \mid D)$ of the model in Equation \eqref{eq:2}; hence, ML was not approximated in this case.

\begin{table}[H]
    \caption{Approximation of ML $\times 10^{-6}$ and weights (\%, in parentheses) of the model for Dataset 1. All results $< 10^{-9}$ are listed as 0. All results $> 10^{-3}$ are listed as ``+''. ``-'' indicates that ML is not calculated.}
    \label{tb:result4}
    \begin{tabular}{>{\centering\arraybackslash}m{0.23\textwidth-2\tabcolsep}>{\centering\arraybackslash}m{0.13\textwidth-2\tabcolsep}>{\centering\arraybackslash}m{0.125\textwidth-2\tabcolsep}>{\centering\arraybackslash}m{0.125\textwidth-2\tabcolsep}>{\centering\arraybackslash}m{0.13\textwidth-2\tabcolsep}>{\centering\arraybackslash}m{0.13\textwidth-2\tabcolsep}>{\centering\arraybackslash}m{0.13\textwidth-2\tabcolsep}}
    \hline
     \multicolumn{7}{c}{BBMD non-informative prior} \\ \hline
     Model & Reference value & MLE-based Schwarz criterion & MCMC-based Schwarz criterion & Laplace approximation & Density estimation & Bridge sampling \\ \hline
    Logistic & 32.5 (24) & + (26) & - (37) & 31.2 (30) & 32.5 (25) & 32.5 (24) \\
    Probit & 9.55 (7) & + (27) & - (41) & 9.50 (9) & 9.49 (7) & 9.53 (7) \\
    Q-linear & 0 (0) & 0.43 (0) & - (0) & 0 (0) & 0 (0) & 0 (0) \\
    Weibull & 4.52 (3) & + (14) & - (7) & 63.9 (61) & 4.16 (3) & 4.49 (3) \\
    Multi-stage2 & 0.01 (0) & 262 (1) & - (0) & 0 (0) & 0.01 (0) & 0.01 (0) \\
    Log Logistic & 55.9 (41) & + (13) & - (7) & 0 (0) & 60.9 (47) & 56.0 (42) \\
    Log Probit & 26.5 (20) & + (14) & - (6) & 0 (0) & 21.0 (16) & 26.3 (20) \\
    Dichotomous-Hill & 5.69 (4) & + (7) & - (2) & 0 (0) & 0.82 (1) & 5.79 (4) \\ \hline
    \multicolumn{7}{c}{\texttt{ToxicR} informative prior} \\ \hline
    Model & Reference value & MLE-based  Schwarz criterion & MCMC-based Schwarz criterion & Laplace approximation & Density estimation & Bridge sampling \\ \hline
    Logistic & 0.27 (1) & + (26) & - (11) & 0.27 (1) & 0.26 (2) & 0.27 (1) \\
    Probit & 10.4 (44) & + (27) & - (46) & 10.4 (49) & 10.3 (96) & 10.4 (44) \\
    Q-linear & 0 (0) & 0.43 (0) & - (0) & 0 (0) & 0 (0) & 0 (0) \\
    Weibull & 8.37 (36) & + (14) & - (19) & 6.38 (30) & 0.01 (0) & 8.33 (35) \\
    Multi-stage2 & 1.46 (6) & 262 (1) & - (0) & 1.34 (6) & 0.04 (0) & 1.46 (6) \\
    Log Logistic & 0.05 (0) & + (13) & - (2) & 0.05 (0) & 0 (0) & 0.05 (0) \\
    Log Probit & 2.58 (11) & + (14) & - (14) & 2.48 (11) & 0.08 (0) & 2.58 (11) \\
    Dichotomous-Hill & 0.37 (2) & + (7) & - (8) & 0.33 (2) & 0 (0) & 0.37 (2) \\ \hline
    \end{tabular}
\end{table}

\begin{table}[H]
    \caption{Approximation of ML $\times 10^{-6}$ and weights (\%, in parentheses) of the model for Dataset 1. All results $< 10^{-9}$ are listed as 0. All results $> 10^{-3}$ are listed as ``+''. ``-'' indicates that ML is not calculated. ``$-^*$'' indicates non-computable value.}
    \label{tb:result5}
    \begin{tabular}{>{\centering\arraybackslash}m{0.23\textwidth-2\tabcolsep}>{\centering\arraybackslash}m{0.13\textwidth-2\tabcolsep}>{\centering\arraybackslash}m{0.125\textwidth-2\tabcolsep}>{\centering\arraybackslash}m{0.125\textwidth-2\tabcolsep}>{\centering\arraybackslash}m{0.13\textwidth-2\tabcolsep}>{\centering\arraybackslash}m{0.13\textwidth-2\tabcolsep}>{\centering\arraybackslash}m{0.13\textwidth-2\tabcolsep}}
    \hline
     \multicolumn{7}{c}{BBMD non-informative prior} \\ \hline
      Model & Reference value & MLE-based  Schwarz criterion & MCMC-based Schwarz criterion & Laplace approximation & Density estimation & Bridge sampling \\ \hline
    Logistic & 0.55 (1) & 875 (20) & - (26) & 0.54 (2) & 0.56 (3) & 0.56 (1) \\
    Probit & 0.18 (0) & 910 (21) & - (27) & 0.18 (1) & 0.18 (1) & 0.18 (0) \\
    Q-linear & 0.64 (1) & 449 (10) & - (14) & 0.62 (3) & 0.64 (3) & 0.63 (1) \\
    Weibull & 3.28 (7) & 474 (11) & - (4) & $-^*$ (0) & 3.49 (17) & 3.29 (7) \\
    Multi-stage2 & 0 (0) & 469 (11) & - (13) & 0.02 (0) & 0.01 (0) & 0 (0) \\
    Log Logistic & 7.12 (15) & 481 (11) & - (7) & 2.73 (12) & 6.47 (32) & 7.10 (15) \\
    Log Probit & 3.65 (8) & 481 (11) & - (5) & 0.97 (4) & 2.34 (11) & 3.68 (8) \\
    Dichotomous-Hill & 31.8 (67) & 241 (5) & - (4) & 17.0 (76) & 6.65 (33) & 31.6 (67) \\\hline
    \multicolumn{7}{c}{\texttt{ToxicR} informative prior} \\ \hline
    Model & Reference value & MLE-based  Schwarz criterion & MCMC-based Schwarz criterion & Laplace approximation & Density estimation & Bridge sampling \\ \hline
    Logistic & 6.97 (2) & 875 (20) & - (15) & 6.80 (2) & 7.06 (15) & 6.95 (2) \\
    Probit & 24.6 (6) & 910 (21) & - (23) & 24.4 (7) & 26.4 (56) & 24.6 (6) \\
    Q-linear & 49.0 (13) & 449 (10) & - (14) & 47.6 (14) & 2.44 (5) & 49.1 (13) \\
    Weibull & 62.5 (16) & 474 (11) & - (11) & 56.2 (16) & 1.38 (3) & 62.5 (16) \\
    Multi-stage2 & 14.3 (4) & 469 (11) & - (10) & 14.6 (4) & 0.56 (1) & 14.2 (4) \\
    Log Logistic & 124 (33) & 481 (11) & - (12) & 115 (32) & 4.99 (11) & 124 (33) \\
    Log Probit & 71.6 (19) & 481 (11) & - (10) & 61.2 (18) & 4.49 (9) & 71.5 (19) \\
    Dichotomous-Hill & 28.3 (7) & 241 (5) & - (5) & 23.5 (7) & 0 (0) & 28.2 (7) \\\hline
    \end{tabular}
\end{table}

The approximations obtained using the MLE- and MCMC-based Schwarz criterion were larger than the reference values, regardless of the prior distribution. The approximation biases obtained by the Laplace approximation were minimal regardless of the prior distribution, except for the Weibull model with a response at zero dosage, as shown in Dataset 4. Regardless of the dataset, the approximation biases were minimal when the \texttt{ToxicR} informative prior was used. The ML of the model with an intercept could not be calculated, or the approximation bias was pronounced in the absence of a zero dosage response, as in Dataset 1, when a prior distribution other than the \texttt{ToxicR} informative prior was used. The approximation biases obtained using the density estimation were minor. However, the values were higher for models with more than two parameters. The approximation biases obtained using the bridge sampling were also minimal regardless of the prior distribution.

\subsection{Estimation of BMD and calculation of BMDL}
\begin{figure}[H]
    \centering
    \includegraphics[width=0.9\textwidth]{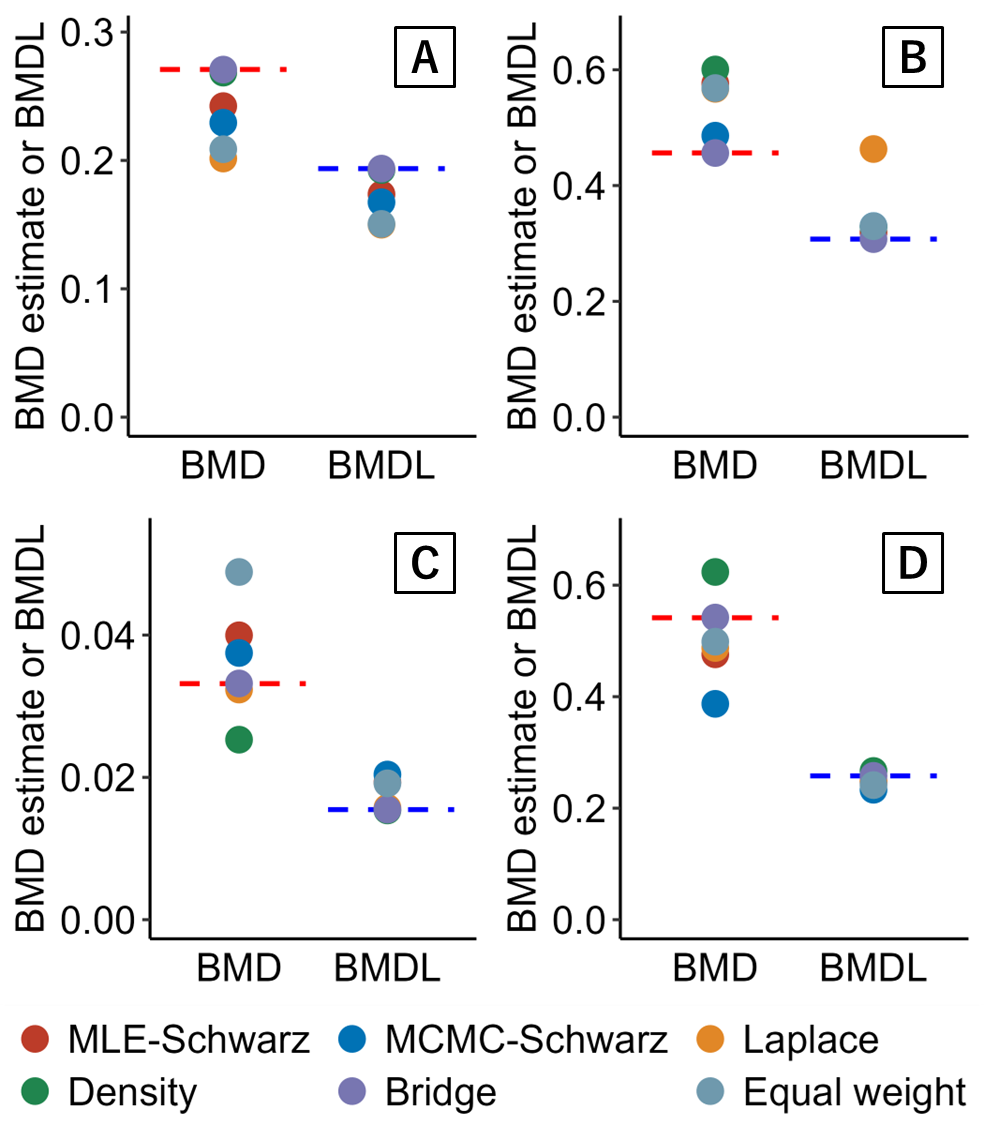}
    \caption{BMD estimates and BMDLs using BBMD non-informative prior. Panel A, Dataset 1; Panel B, Dataset 2; Panel C, Dataset 3; and Panel D, Dataset 4. The red-dashed line is the BMD reference value, and the blue-dashed line is the BMDL reference value. MLE: Schwarz criterion using the maximum likelihood; MCMC: Markov chain Monte Carlo.}
    \label{fig:2}
\end{figure}

\begin{figure}[H]
    \centering
    \includegraphics[width=0.9\textwidth]{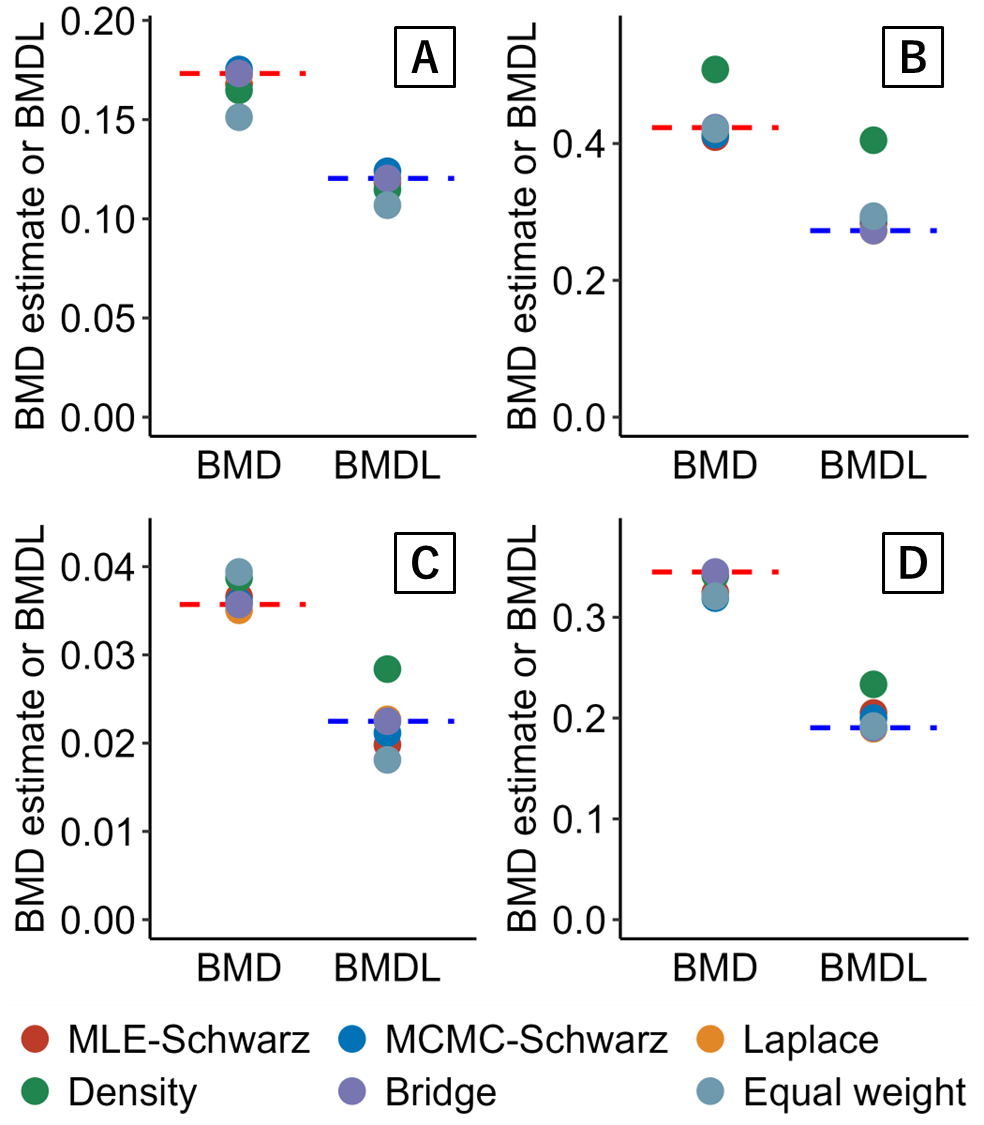}
    \caption{BMD estimates and BMDLs using \texttt{ToxicR} informative prior. Panel A, Dataset 1; Panel B, Dataset 2; Panel C, Dataset 3; and Panel D, Dataset 4. The red-dashed line is the BMD reference value, and the blue-dashed line is the BMDL reference value. MLE: Schwarz criterion using the maximum likelihood; MCMC: Markov chain Monte Carlo.}
    \label{fig:3}
\end{figure}

The calculated BMD estimates and BMDLs are shown in Table \ref{tb:result6} and Figures \ref{fig:2} and \ref{fig:3}. The BMD estimate and BMDL results (equal weights), without considering the weights of the BMD posterior distribution, were included to evaluate the effects after considering the weights of the BMD posterior distribution. The results obtained using the informative prior based on previous data are provided in Section S3.2 of the Supplementary materials.

\begin{table}[H]
    \caption{BMD estimates and BMDL values (in parentheses). Upper table: Results on applying BBMD non-informative prior; lower table: results on applying \texttt{ToxicR} informative prior.}
    \label{tb:result6}
    \begin{tabular}{>{\centering\arraybackslash}m{0.1\textwidth-2\tabcolsep}>{\centering\arraybackslash}m{0.13\textwidth-2\tabcolsep}>{\centering\arraybackslash}m{0.13\textwidth-2\tabcolsep}>{\centering\arraybackslash}m{0.13\textwidth-2\tabcolsep}>{\centering\arraybackslash}m{0.13\textwidth-2\tabcolsep}>{\centering\arraybackslash}m{0.13\textwidth-2\tabcolsep}>{\centering\arraybackslash}m{0.13\textwidth-2\tabcolsep}>{\centering\arraybackslash}m{0.12\textwidth-2\tabcolsep}}
     \multicolumn{8}{c}{BBMD non-informative prior} \\ \hline
     Dataset & Reference value & MLE-based Schwarz criterion & MCMC-based Schwarz  criterion & Laplace approximation & Density estimation & Bridge sampling & Equal weight \\ \hline
     1 & 0.270\par(0.194) & 0.243\par(0.174) & 0.229\par(0.167) & 0.201\par(0.150) & 0.268\par(0.193) & 0.271\par(0.194) & 0.209\par(0.151) \\
    2 & 0.456\par(0.308) & 0.573\par(0.319) & 0.486\par(0.330) & 0.567\par(0.463) & 0.601\par(0.329) & 0.456\par(0.308) & 0.568\par(0.330) \\
    3 & 0.033\par(0.015) & 0.040\par(0.019) & 0.037\par(0.020) & 0.032\par(0.016) & 0.025\par(0.015) & 0.033\par(0.015) & 0.049\par(0.019) \\
    4 & 0.541\par(0.258) & 0.476\par(0.247) & 0.387\par(0.233) & 0.487\par(0.249) & 0.624\par(0.267) & 0.542\par(0.258) & 0.499\par(0.241) \\ \hline
    \multicolumn{8}{c}{\texttt{ToxicR} informative prior} \\ \hline
    Dataset & Reference value & MLE-based Schwarz criterion & MCMC-based Schwarz  criterion & Laplace approximation & Density estimation & Bridge sampling & Equal weight \\ \hline
    1 & 0.173\par(0.120) & 0.169\par(0.119) & 0.175\par(0.124) & 0.172\par(0.120) & 0.165\par(0.115) & 0.173\par(0.120) & 0.151\par(0.107) \\
    2 & 0.423\par(0.273) & 0.409\par(0.283) & 0.413\par(0.291) & 0.421\par(0.272) & 0.508\par(0.404) & 0.423\par(0.273) & 0.422\par(0.293) \\
    3 & 0.036\par(0.022) & 0.037\par(0.020) & 0.036\par(0.021) & 0.035\par(0.023) & 0.039\par(0.028) & 0.036\par(0.022) & 0.039\par(0.018) \\
    4 & 0.345\par(0.190) & 0.324\par(0.205) & 0.319\par(0.200) & 0.342\par(0.189) & 0.342\par(0.223) & 0.345\par(0.190) & 0.321\par(0.191) \\ \hline
    \end{tabular}
\end{table}

First, we considered the BMD estimates for the approximation results using BBMD non-informative prior (except for equal weights), that is, the upper part of Table \ref{tb:result6}. For Dataset 1, the bias in BMD obtained using the Laplace approximation ($6.94 \times 10^{-2}$) was the largest. For Datasets 2 and 3, the biases in BMD obtained using the density estimation ($1.44 \times 10^{-1}$ and $7.88 \times 10^{-3}$, respectively) were the largest. For Dataset 4, the bias in BMD obtained using the MCMC-based Schwarz criterion ($1.54 \times 10^{-1}$) was the largest. The biases in BMD obtained using the bridge sampling were small, regardless of the dataset. These characteristics were similar to those of the BMDL.

Second, we considered the BMD estimates obtained when \texttt{ToxicR} informative prior was used (except for equal weights), that is, the lower part of Table \ref{tb:result6}. For Datasets 1–3, the biases in BMD obtained using the density estimation ($8.23\times10^{-3}, 8.50\times10^{-2}, 3.02\times10^{-3}$, respectively) were the largest. For Dataset 4, the biases in BMD obtained using the MCMC-based Schwarz criterion estimate ($2.59\times10^{-2}$) were the largest. The biases in the BMD obtained using the bridge sampling were small regardless of the dataset. These characteristics were similar to those of the BMDL.

Overall, the bias in BMD obtained using the MLE-based Schwarz criterion, MCMC-based Schwarz criterion, or the density estimation tended to be large.

The biases in the BMD obtained using the Laplace approximation were small when the \texttt{ToxicR} informative prior was used. The BMD biases obtained using the bridge sampling were almost equal to the reference values. For all datasets, the biases in the BMD obtained with equal weights tend to be large. Therefore, it was confirmed that the weights of the models should be considered.

\section{Discussion}
We discuss the reasons for the large biases (except when bridge sampling was used) when computing the approximated ML value. The approximated ML values obtained using the MLE- and MCMC-based Schwarz criterion (used in BBMD) were larger than the reference values. This is because the Schwarz criterion does not consider the prior or posterior distributions (see Equation \eqref{eq:5}), resulting in an inadequate approximation of the ML for the assumed prior distribution.

In the Laplace approximation (used in \texttt{ToxicR}) with the BBMD non-informative prior for Datasets 1 and 2, the MAP estimate of the intercept parameter was close to zero, which was the lower limit of the parameter. Thus, in computing the numerical derivative to find the negative Hessian matrix, the determinant and log-transformed value were not calculated appropriately because the width of the numerical derivative was too narrow, and the ML could not be calculated. Even when the determinant and log-transformed value were calculated, the approximation results were far from the reference values. For Datasets 3 and 4, the above problem did not arise because the MAP estimate (except for parameter $\alpha$ in the Weibull model) was far from zero. Similarly, this problem did not arise when the \texttt{ToxicR} informative prior was used because the MAP estimate was slightly different from zero, even for datasets with zero response at zero dosage.

The approximation biases obtained using the density estimation tended to be minimal. However, the biases tended to be large for models with more than two parameters. This is because the number of calculation points for the parameter decreases as the number of parameters increases in the default setting of the \texttt{ks} package. Although the number of points could be increased, an enormous amount of time would be required. When the \texttt{ToxicR} informative prior was used, the approximation biases were large for models with the intercept parameter; additionally, the posterior distribution was concentrated around one point and the value of the posterior distribution at the estimated parameter was extremely large. Therefore, the denominator in Equation \eqref{eq:6} increased, resulting in a larger approximation bias.

The computation times for each dataset were 2-3 hours for the reference value; 1-2 seconds for the MLE- and MCMC-based Schwarz criterion, and the Laplace approximation; approximately 3 minutes for the density estimation; and approximately 30 seconds for the bridge sampling. 

Next, we discuss the impact of the approximation methods on the BMD estimates and BMDL calculations. For Dataset 4, the approximation bias obtained using the density estimation with the BBMD non-informative prior was small. However, the posterior probability based on the approximated ML value was far from that based on the reference value, and the BMD estimate and BMDL were far from the respective reference values.

Thus, we presume that the small difference between the approximate ML value and reference value strongly affected the BMD estimate and BMDL calculation. The impact of the ML approximation method on the BMD estimate was minimal for Dataset 3. We believe that the differences in BMD estimates and BMDL among models were small because the response occurred at lower doses in Dataset 3, and the shapes of the model-fitting curves were similar.

Based on our observations, we do not recommend the MLE- or MCMC-based Schwarz criterion because the approximation biases and the differences between the posterior probabilities of the approximated ML values and the reference values were large. In the Laplace approximation, ML was not calculated when the MAP estimator of the parameter was small. Thus, the prior distribution should be selected carefully when using the Laplace approximation. We do not recommend the density estimation because the approximation biases were large for models with three or more parameters, and the calculation cost was large. The approximation bias obtained using the bridge sampling was small because the bridge sampling was a direct improvement over naive Monte Carlo integration.

Bayesian BMD has the advantage of being able to use the bridge sampling, producing better results than those by using other ML approximation methods even for arbitrary prior distribution settings. Therefore, we recommend using Bayesian BMD or implementing bridge sampling in BBMD and \texttt{ToxicR} for better BMD estimation.

This study has two limitations. First, we used the four datasets but did not consider the true BMD. Therefore, the bias in BMD in this study is not a true bias. In the future, a simulation study should be performed using the BMD of true models to investigate whether ML approximation methods affect BMD estimation. Second, the settings of models and prior distributions implemented in Bayesian BMD were not considered in this study for simplicity of interpretation. Future research should investigate these settings and verify whether the results obtained are the same as those in this study.

\section{Conclusion}
Several ML approximation methods are employed in standard software packages, such as BBMD, \texttt{ToxicR}, and Bayesian BMD for the BMD method, because the analytical calculation of ML is challenging. Although ML values differ among the approximation methods, resulting in different posterior probabilities and BMD estimates, the methods are neither widely recognized nor quantitatively evaluated.  

In this study, we evaluated the performance of five ML approximation methods. The approximation and estimation biases of bridge sampling were the smallest regardless of the dataset or prior distributions. Both the approximation and estimation biases of the MCMC-based Schwarz criterion and Laplace approximation were large for some datasets. Thus, the approximation biases of the density estimation were relatively small for some datasets but large for others. In terms of the accuracy of ML approximation methods, using Bayesian BMD with bridge sampling is preferred.

\section*{Data statement}
All the results in Section 3 were given in the Supplementary materials.

\section*{Acknowledgements}
We would like to thank Editage (www.editage.jp) for English language editing.

\section*{Funding}
 The authors declare that no grants, funds, or other support were received during the preparation of this article.

\section*{Author contributions}
\textbf{Sota Minewaki}: Conceptualization, Methodology, Software, Formal analysis, Investigation, Writing – original draft, Writing – review \& editing, Visualization. \textbf{Tomohiro Ohigashi}: Writing – review \& editing, Supervision. \textbf{Takashi Sozu}: Conceptualization, Writing – review \& editing, Supervision.

\end{document}


\maketitle
\section{Informative prior based on previous data}
In this section, we explain the specification of the informative prior based on previous data (hereafter referred to as the data-based informative prior) \cite{r3,r4,r5}.
The data-based informative prior is a prior distribution based on 518 datasets used in Shao and Shapiro (2018) \cite{r3}. In this study, we derived the data-based informative prior by following the same approach as the study.

To obtain the data-based informative prior, hierarchical structures were set for the model parameters of each model, as shown in Table \ref{tb:3}. 
The prior and hyperprior distributions for the model parameters followed the same distributions as those used in Shao and Shapiro (2018).
The parameter values of the hyperprior distributions were configured to make the prior as uninformative as possible. Since Shao and Shapiro (2018) did not disclose the parameter values of the hyperprior distributions, we set these values as shown in Table \ref{tb:3}.

\begin{table}[H]
    \caption{Prior and hyper distribution for data based informative prior}
    \label{tb:3}
    \renewcommand{\arraystretch}{1.4}
    \begin{tabular}{>{\centering\arraybackslash}m{0.255\textwidth-2\tabcolsep}>{\centering\arraybackslash}m{0.179\textwidth-2\tabcolsep}>{\centering\arraybackslash}m{0.349\textwidth-2\tabcolsep}>{\centering\arraybackslash}m{0.221\textwidth-2\tabcolsep}}
    \hline
        Models & Prior distribution & Hyperprior distribution & Data based informative prior \\ \hline
        Logistic : $\frac{1}{1 + \exp{(-\alpha - \beta d)}}$& $\alpha \sim \text{N}(\mu_\alpha, \sigma_\alpha)$ $\beta \sim \text{Ga}(s_\beta, r_\beta)$ & $\mu_\alpha \sim \text{U}(-10,10)$, $\sigma_\alpha \sim \text{U}(0,50)$ $s_\beta \sim \text{U}(0,15)$, $r_\beta \sim \text{U}(0,10)$ & $\alpha \sim \text{N}(-2.79, 1.49)$ $\beta \sim \text{Ga}(2.01, 0.49)$ \\ \hline
        Probit : $\Phi(\alpha + \beta d)$ & $\alpha \sim \text{N}(\mu_\alpha, \sigma_\alpha)$ $\beta \sim \text{Ga}(s_\beta, r_\beta)$ & $\mu_\alpha \sim \text{U}(-10,10)$ $\sigma_\alpha \sim \text{U}(0,50)$ $s_\beta \sim \text{U}(0,15)$ $r_\beta \sim \text{U}(0,10)$ & $\alpha \sim \text{N}(-1.57, 0.80)$ $\beta \sim \text{Ga}(2.07, 0.90)$ \\ \hline
        Q-linear : $\beta + (1-\beta)$\par$(1-\exp{(-\alpha d)})$ & $\alpha \sim \text{Ga}(s_\alpha, r_\alpha)$ $\beta \sim \text{Be}(\alpha_\beta, \beta_\beta)$ & $s_\alpha \sim \text{U}(0,15)$, $r_\alpha \sim \text{U}(0,10)$ $\alpha_\beta \sim \text{U}(0,10)$, $\beta_\beta \sim \text{U}(0,10)$ & $\alpha \sim \text{Ga}(0.92, 0.74)$ $\beta \sim \text{Be}(0.29, 3.26)$ \\ \hline
        Weibull : $\gamma + (1-\gamma)$\newline$(1-\exp{(-\beta\exp{(\alpha)})})$ & $\alpha \sim \text{Ga}(s_\alpha, r_\alpha)$ $\beta \sim \text{Ga}(s_\beta, r_\beta)$  $\gamma \sim \text{Be}(\alpha_\gamma, \beta_\gamma)$ & $s_\alpha \sim \text{U}(0,15)$, $r_\alpha \sim \text{U}(0,10)$  $s_\beta \sim \text{U}(0,10)$, $r_\beta \sim \text{U}(0,10)$ $\alpha_\gamma \sim \text{U}(0,10)$, $\beta_\gamma \sim \text{U}(0,10)$ & $\alpha \sim \text{Ga}(2.55, 1.60)$ $\beta \sim \text{Ga}(0.87, 0.60)$ $\gamma \sim \text{Be}(0.31, 3.64)$ \\ \hline
        Multi-stage2 : $\gamma + (1-\gamma)$\newline$(1-\exp{(-\alpha d -\beta d^2)})$ & $\alpha \sim \text{Ga}(s_\alpha, r_\alpha)$ $\beta \sim \text{Ga}(s_\beta, r_\beta)$  $\gamma \sim \text{Be}(\alpha_\gamma, \beta_\gamma)$ & $s_\alpha \sim \text{U}(0,15)$, $r_\alpha \sim \text{U}(0,10)$  $s_\beta \sim \text{U}(0,10)$, $r_\beta \sim \text{U}(0,10)$ $\alpha_\gamma \sim \text{U}(0,10)$, $\beta_\gamma \sim \text{U}(0,10)$ & $\alpha \sim \text{Ga}(0.29, 0.30)$ $\beta \sim \text{Ga}(0.48, 0.86)$ $\gamma \sim \text{Be}(0.31, 3.33)$ \\ \hline
        Log Logistic : $\gamma + \frac{1-\gamma}{1+\exp{(-\alpha - \beta \log{d}})}$ & $\alpha \sim \text{N}(\mu_\alpha, \sigma_\alpha)$ $\beta \sim \text{Ga}(s_\beta, r_\beta)$  $\gamma \sim \text{Be}(\alpha_\gamma, \beta_\gamma)$ & $\mu_\alpha \sim \text{U}(-10,10)$, $\sigma_\alpha \sim \text{U}(0,50)$  $s_\beta \sim \text{U}(0,10)$, $r_\beta \sim \text{U}(0,10)$ $\alpha_\gamma \sim \text{U}(0,10)$, $\beta_\gamma \sim \text{U}(0,10)$ & $\alpha \sim \text{N}(0.41, 1.92)$ $\beta \sim \text{Ga}(2.72, 1.24)$ $\gamma \sim \text{Be}(0.32, 3.66)$ \\ \hline
        Log Probit : $\gamma + (1-\gamma)\Phi(\alpha + \beta \log{d})$ & $\alpha \sim \text{N}(\mu_\alpha, \sigma_\alpha)$ $\beta \sim \text{Ga}(s_\beta, r_\beta)$  $\gamma \sim \text{Be}(\alpha_\gamma, \beta_\gamma)$ & $\mu_\alpha \sim \text{U}(-10,10)$, $\sigma_\alpha \sim \text{U}(0,50)$  $s_\beta \sim \text{U}(0,10)$, $r_\beta \sim \text{U}(0,10)$ $\alpha_\gamma \sim \text{U}(0,10)$, $\beta_\gamma \sim \text{U}(0,10)$ & $\alpha \sim \text{N}(0.23, 1.11)$ $\beta \sim \text{Ga}(2.63, 2.12)$ $\gamma \sim \text{Be}(0.33, 3.83)$ \\ \hline
        Dichotomous-Hill : $\delta\left( \gamma + \frac{1-\gamma}{1+\exp{(-\alpha - \beta \log{d}})} \right)$ & $\alpha \sim \text{N}(\mu_\alpha, \sigma_\alpha)$ $\beta \sim \text{Ga}(s_\beta, r_\beta)$  $\gamma \sim \text{Be}(\alpha_\gamma, \beta_\gamma)$ $\delta \sim \text{Be}(\alpha_\delta, \beta_\delta)$ & $\mu_\alpha \sim \text{U}(-10,10)$, $\sigma_\alpha \sim \text{U}(0,50)$  $s_\beta \sim \text{U}(0,10)$, $r_\beta \sim \text{U}(0,10)$ $\alpha_\gamma \sim \text{U}(0,10)$, $\beta_\gamma \sim \text{U}(0,10)$ $\alpha_\delta \sim \text{U}(0,10)$, $\beta_\delta \sim \text{U}(0,10)$ & $\alpha \sim \text{N}(0.23, 1.11)$ $\beta \sim \text{Ga}(4.46, 1.52)$ $\gamma \sim \text{Be}(0.35, 2.78)$ $\delta \sim \text{Be}(1.18, 0.39)$ \\ \hline
    \end{tabular}
\end{table}

Using this hierarchical structure, MCMC was conducted with \texttt{rstan} on 518 datasets.
To simplify the explanation of the process for deriving data-based informative priors, we focus on the $\alpha$ parameter of the logistic model. 
Through MCMC, posterior samples of the hyperparameters $\mu_\alpha^{'}$ and $\sigma_\alpha^{'}$ are obtained. 
These posterior samples are subsequently used to draw $\hat{\alpha}$ from the predictive distribution $\hat{\alpha} \sim \text{N}(\mu_\alpha^{'}, \sigma_\alpha^{'})$. 
By applying the method of moments to the sampled $\hat{\alpha}$, the parameters of the prior distribution for $\alpha$ are estimated. 
The prior distribution obtained in this way is the data-based informative prior, as shown in Table \ref{tb:3}.

\section{Bayesian BMD}
Bayesian BMD is newly released web software developed by EFSA \cite{r1,r2}. The available models and prior distributions differ from those implemented in other software and used in previous studies on BMD method using BMA. In this section, we explain the models and prior distributions used in Bayesian BMD.

\subsection{Models}
In Bayesian BMD, eight models are available for dichotomous datasets. These models differ from those used in BBMD and ToxicR. The models used in Bayesian BMD are shown in Table \ref{tb:1}

\begin{table}[H]
    \caption{Models used in Bayesian BMD}
    \label{tb:1}
    \renewcommand{\arraystretch}{1.4}
    \begin{tabular}{>{\centering\arraybackslash}m{0.7\textwidth-2\tabcolsep}>{\centering\arraybackslash}m{0.3\textwidth-2\tabcolsep}}
    \hline
        Models & Constraints \\ \hline
        Exponential : $a + (1 - a)(1 - \exp{(-bd^c)})$& \multirow{8}{*}{$0 \leq a \leq 1$, $0 \leq b,c$} \\ \cline{1-1}
        Inverse exponential : $a + (1-a)\exp{(-bd^c)}$ &   \\ \cline{1-1}
        Hill : $a + (1-a)(1-\frac{b}{b+d^c})$ &  \\ \cline{1-1}
        Log normal : $a + (1-a)\Phi(\log(b) + c\log(d))$ &  \\ \cline{1-1}
        Gamma : $a + (1-a)\frac{\gamma(c,bd)}{\Gamma(c)}$ &  \\ \cline{1-1}
        Quadratic exponential : $a + (1-a)(1-\exp{(-bd-cd^2)})$ &  \\ \cline{1-1}
        Probit increasing : $\Phi(\Phi^{-1}(a)+bd^c)$ &  \\ \cline{1-1}
        Logistic increasing : $\text{expit}(\text{logit}(a)+bd^c)$ &  \\ \hline
    \end{tabular}
    \caption*{$d$ is the dose, $\gamma(s,v)$ is lower incomplete gamma function, expressed as $\gamma(s,v) = \int^v_0 u^{s-1}\exp{(-u)}du$, and $\Gamma(\cdot)$ is gamma function. $\Phi(\cdot)$ is  the cumulative standard normal distribution and $\Phi^-1(\cdot)$ is its inverse function. $\text{expit}(x) = \frac{\exp{(\pi/\sqrt{3})x}}{1+\exp{(\pi/\sqrt{3})}x}$, and $\text{logit}(x) = \frac{\sqrt{3}}{\pi}\log{(\frac{x}{1-x})}$.}
\end{table}

\subsection{Prior distributions}

\subsubsection{PERT prior distribution}
The two-parameter beta distribution is defined for $x \in [0,1]$, with parameters $\alpha$ and $\beta$. The probability density function (PDF) of the four-parameter beta distribution, defined for $y \in [l,u]$ using the transformation $y = x(u-l) + l$, is expressed as follows:

\begin{equation}
\label{eq:1}
f(y \mid \alpha, \beta, l, u) = \frac{\text{Be}(x \mid \alpha, \beta)}{u-l} = \frac{(y-l)^{\alpha -1}(u-y)^{\beta- 1}}{(u-l)^{\alpha + \beta - 1}\text{B}(\alpha, \beta)}, 
\end{equation}

\noindent
where $\text{Be}(\cdot \mid \alpha, \beta)$ denotes the beta distribution and $\text{B}(\alpha, \beta)$ denotes the beta function. The distribution that expresses $\alpha$ and $\beta$ in Equation (\ref{eq:1}) using following equations is program evaluation and review technique (PERT) distribution:

$$
\alpha = \frac{4m + u - 5l}{u-l},
$$
$$
\beta = \frac{5u -l - 4m}{u-l},
$$

\noindent
where $m$ denotes the mode, $l$ represents the lower bound, and $u$ represents the upper bound. The flatness of the PERT distribution can be controlled by introducing an additional parameter $\gamma$. This results in the modified PERT distribution, with the following parameters:

$$
\alpha = 1 + \gamma \frac{m-l}{u-l},
$$
$$
\beta = 1 + \gamma \frac{u -m}{u-l}.
$$

\noindent
The parameter $\gamma$ takes positive real values, with smaller values resulting in a flatter distribution.
When $\gamma = 0.0001$, the distribution becomes a uniform distribution on $[l,u]$.

\subsubsection{Default prior distribution}
In Bayesian BMD, $a$ and $b$ parameters of each model in Table \ref{tb:1} are transformed into natural parameters (background response, BMD, and maximal response), which are easier for users to interpret, and then the prior distributions are set. The background response is expressed as $f(0)$, where $f(d)$ represents the model value at dose $d$. For dichotomous datasets, only the background response and BMD are used because the maximal response is equal to 1. 
Modified PERT distributions are employed for the prior distributions of natural parameters.
The technical parameter $c$ associated with each model in Table \ref{tb:1}, is log-transformed prior to specifying a normal distribution as the prior.
The default prior distributions in Bayesian BMD are summarized in Table \ref{tb:2}.

\begin{table}[H]
    \caption{The setting of default prior distributions of Bayesian BMD}
    \label{tb:2}
    \renewcommand{\arraystretch}{1.4}
    \begin{tabular}{>{\centering\arraybackslash}m{0.33\textwidth-2\tabcolsep}>{\centering\arraybackslash}m{0.33\textwidth-2\tabcolsep}>{\centering\arraybackslash}m{0.33\textwidth-2\tabcolsep}}
    \hline
        Background response & BMD & d \\ \hline
        $l = \max(\text{LCL}_1, \frac{1}{10n_1})$ & $l = 0$ & \multirow{4}{*}{$\log(d) \sim \text{N}(0,1^2)$} \\ \cline{1-2}
        $m = \max(\bar{p}_1, \frac{1}{5n_1})$ & $m = 0.5$ & \\ \cline{1-2}
        $u = \min(\text{UCL}_1, 1-\frac{1}{10n_1})$ & $u = (\max d)^2$ & \\ \cline{1-2}
        $\gamma = 4$ & $\gamma = 0.0001$ \\ \hline        
    \end{tabular}
    \caption*{$\bar{p}_1$ is the observed proportion in the control dose group, $n_1$ is the size of background, and $\text{LCL}_1$, $\text{UCL}_1$ are the confidence limits of $\bar{p}_1$ based on a binomial test with $\text{H}_0: p_1 = 0.5$. $u = \max d$ is the highest dose level.}
\end{table}

\section{Result}
\subsection{ML estimates and weights of the model}
\begin{table}[H]
    \caption{Approximation of ML $\times 10^{-6}$ and weights (\%, in parentheses) of the model for Dataset 1. All results $< 10^{-9}$ are listed as 0. All results $> 10^{-3}$ are listed as ``+''. ``-'' indicates that ML is not calculated. ``$-^*$'' indicates non-computable value.}
    \label{tb:4}
    \begin{tabular}{>{\centering\arraybackslash}m{0.23\textwidth-2\tabcolsep}>{\centering\arraybackslash}m{0.13\textwidth-2\tabcolsep}>{\centering\arraybackslash}m{0.125\textwidth-2\tabcolsep}>{\centering\arraybackslash}m{0.125\textwidth-2\tabcolsep}>{\centering\arraybackslash}m{0.13\textwidth-2\tabcolsep}>{\centering\arraybackslash}m{0.13\textwidth-2\tabcolsep}>{\centering\arraybackslash}m{0.13\textwidth-2\tabcolsep}}
    \hline
     \multicolumn{7}{c}{BBMD non-informative prior} \\ \hline
      Model & Reference value & MLE-based  Schwarz criterion & MCMC-based Schwarz criterion & Laplace approximation & Density estimation & Bridge sampling \\ \hline
    Logistic & 32.5(24) & +(26) & -(37) & 31.2(30) & 32.5(25) & 32.5(24) \\ 
    Probit & 9.55(7) & +(27) & -(41) & 9.50(9) & 9.49(7) & 9.53(7) \\ 
    Q-linear & 0(0) & 0.43(0) & -(0) & 0(0) & 0(0) & 0(0) \\ 
    Weibull & 4.52(3) & +(14) & -(7) & 63.9(61) & 4.16(3) & 4.49(3) \\ 
    Multi-stage2 & 0.01(0) & 262(1) & -(0) & 0(0) & 0.01(0) & 0.01(0) \\ 
    Log Logistic & 55.9(41) & +(13) & -(7) & 0(0) & 60.9(47) & 56.0(42) \\
    Log Probit & 26.5(20) & +(14) & -(6) & 0(0) & 21.0(16) & 26.3(20) \\
    Dichotomous-Hill & 5.69(4) & +(7) & -(2) & 0(0) & 0.82(1) & 5.79(4) \\ \hline
    \multicolumn{7}{c}{\texttt{ToxicR} informative prior} \\ \hline
    Logistic & 0.27(1) & +(26) & -(11) & 0.27(1) & 0.26(2) & 0.27(1) \\
    Probit & 10.4(44) & +(27) & -(46) & 10.4(49) & 10.3(96) & 10.4(44) \\
    Q-linear & 0(0) & 0.43(0) & -(0) & 0(0) & 0(0) & 0(0) \\
    Weibull & 8.37(36) & +(14) & -(19) & 6.38(30) & 0.01(0) & 8.33(35) \\
    Multi-stage2 & 1.46(6) & 262(1) & -(0) & 1.34(6) & 0.04(0) & 1.46(6) \\
    Log Logistic & 0.05(0) & +(13) & -(2) & 0.05(0) & 0(0) & 0.05(0) \\
    Log Probit & 2.58(11) & +(14) & -(14) & 2.48(11) & 0.08(0) & 2.58(11) \\
    Dichotomous-Hill & 0.37(2) & +(7) & -(8) & 0.33(2) & 0(0) & 0.37(2) \\ \hline
        \multicolumn{7}{c}{Data based informative prior} \\ \hline
    Logistic & 126(45) & +(26) & -(35) & 0(0) & 132(54) & 126(45) \\
    Probit & 88.3(32) & +(27) & -(30) & 0.31(100) & 88.7(36) & 88.3(32) \\
    Q-linear & 0.09(0) & 0.43(0) & -(0) & $-^*$ (0) & 0.09(0) & 0.09(0) \\
    Weibull & 11.4(4) & +(14) & -(5) & $-^*$ (0) & 5.33(2) & 11.4(4) \\
    Multi-stage2 & 3.24(1) & 262(1) & -(1) & $-^*$ (0) & 0.79(0) & 3.22(1) \\
    Log Logistic & 28.0(10) & +(13) & -(14) & $-^*$ (0) & 11.2(5) & 28.1(10) \\
    Log Probit & 21.7(8) & +(14) & -(13) & $-^*$ (0) & 7.81(3) & 21.7(8) \\
    Dichotomous-Hill & 0.49(2) & +(7) & -(2) & $-^*$ (0) & 0(0) & 0.51(2) \\ \hline
    \end{tabular}
\end{table}

\begin{table}[H]
    \caption{Approximation of ML $\times 10^{-6}$ and weights (\%, in parentheses) of the model for Dataset 2. All results $< 10^{-9}$ are listed as 0. All results $> 10^{-3}$ are listed as ``+''. ``-'' indicates that ML is not calculated. ``$-^*$'' indicates non-computable value.}
    \label{tb:5}
    \begin{tabular}{>{\centering\arraybackslash}m{0.23\textwidth-2\tabcolsep}>{\centering\arraybackslash}m{0.13\textwidth-2\tabcolsep}>{\centering\arraybackslash}m{0.125\textwidth-2\tabcolsep}>{\centering\arraybackslash}m{0.125\textwidth-2\tabcolsep}>{\centering\arraybackslash}m{0.13\textwidth-2\tabcolsep}>{\centering\arraybackslash}m{0.13\textwidth-2\tabcolsep}>{\centering\arraybackslash}m{0.13\textwidth-2\tabcolsep}}
    \hline
     \multicolumn{7}{c}{BBMD non-informative prior} \\ \hline
      Model & Reference value & MLE-based  Schwarz criterion & MCMC-based Schwarz criterion & Laplace approximation & Density estimation & Bridge sampling \\ \hline
    Logistic & 0.34(3) & 692(9) & -(17) & 0.33(73) & 0.36(9) & 0.34(3) \\
    Probit & 0.12(1) & 915(12) & -(22) & 0.12(27) & 0.12(3) & 0.12(1) \\
    Q-linear & 0.13(1) & +(16) & -(21) & 0(0) & 0.14(4) & 0.14(1) \\
    Weibull & 0.40(3) & +(15) & -(4) & $-^*$ (0) & 0.44(12) & 0.40(3) \\
    Multi-stage2 & 0(0) & +(16) & -(19) & 0(0) & 0(0) & 0(0) \\
    Log Logistic & 1.26(10) & +(15) & -(7) & 0(0) & 1.02(27) & 1.25(10) \\
    Log Probit & 0.59(5) & +(13) & -(5) & $-^*$ (0) & 0.58(15) & 0.59(5) \\
    Dichotomous-Hill & 9.69(77) & 432(5) & -(4) & 0(0) & 1.18(31) & 9.67(77) \\ \hline
    \multicolumn{7}{c}{\texttt{ToxicR} informative prior} \\ \hline
    Logistic & 0.12(0) & 692(9) & -(8) & 0.12(0) & 0.12(2) & 0.12(0) \\
    Probit & 5.18(5) & 915(12) & -(19) & 5.14(6) & 5.66(90) & 5.18(5) \\
    Q-linear & 10.6(11) & +(16) & -(18) & 10.3(13) & 0.10(2) & 10.6(11) \\
    Weibull & 17.3(18) & +(15) & -(14) & 15.7(19) & 0.07(1) & 17.3(18) \\
    Multi-stage2 & 0.40(0) & +(16) & -(12) & 0.40(0) & 0(0) & 0.40(0) \\
    Log Logistic & 29.9(32) & +(15) & -(13) & 26.8(33) & 0.24(4) & 29.9(32) \\
    Log Probit & 12.2(13) & +(13) & -(8) & 8.92(11) & 0.10(2) & 12.2(13) \\
    Dichotomous-Hill & 19.0(20) & 432(5) & -(7) & 14.1(17) & 0(0) & 18.9(20) \\ \hline
        \multicolumn{7}{c}{Data based informative prior} \\ \hline
    Logistic & 89.8(6) & 692(9) & -(10) & 32.0(20) & 92.6(9) & 89.8(6) \\
    Probit & 118(7) & 915(12) & -(13) & 126(80) & 119(11) & 118(7) \\
    Q-linear & 242(15) & +(16) & -(20) & $-^*$ (0) & 271(25) & 242(15) \\
    Weibull & 208(13) & +(15) & -(14) & $-^*$ (0) & 118(11) & 207(13) \\
    Multi-stage2 & 333(21) & +(16) & -(14) & $-^*$ (0) & 223(21) & 338(21) \\
    Log Logistic & 186(12) & +(15) & -(13) & $-^*$ (0) & 176(16) & 187(12) \\
    Log Probit & 152(9) & +(13) & -(11) & $-^*$ (0) & 73.8(7) & 153(10) \\
    Dichotomous-Hill & 270(17) & 432(5) & -(6) & $-^*$ (0) & 1.94(0) & 250(16) \\\hline
    \end{tabular}
\end{table}

\begin{table}[H]
    \caption{Approximation of ML $\times 10^{-6}$ and weights (\%, in parentheses) of the model for Dataset 3. All results $< 10^{-9}$ are listed as 0. All results $> 10^{-3}$ are listed as ``+''. ``-'' indicates that ML is not calculated. ``$-^*$'' indicates non-computable value.}
    \label{tb:6}
    \begin{tabular}{>{\centering\arraybackslash}m{0.23\textwidth-2\tabcolsep}>{\centering\arraybackslash}m{0.13\textwidth-2\tabcolsep}>{\centering\arraybackslash}m{0.125\textwidth-2\tabcolsep}>{\centering\arraybackslash}m{0.125\textwidth-2\tabcolsep}>{\centering\arraybackslash}m{0.13\textwidth-2\tabcolsep}>{\centering\arraybackslash}m{0.13\textwidth-2\tabcolsep}>{\centering\arraybackslash}m{0.13\textwidth-2\tabcolsep}}
    \hline
     \multicolumn{7}{c}{BBMD non-informative prior} \\ \hline
      Model & Reference value & MLE-based  Schwarz criterion & MCMC-based Schwarz criterion & Laplace approximation & Density estimation & Bridge sampling \\ \hline
    Logistic & 0.28(2) & 230(18) & -(21) & 0.27(2) & 0.29(3) & 0.28(2) \\
    Probit & 0.05(0) & 157(12) & -(15) & 0.05(0) & 0.06(1) & 0.05(0) \\
    Q-linear & 7.96(64) & 340(26) & -(30) & 8.13(66) & 8.41(80) & 7.96(64) \\
    Weibull & 1.14(9) & 152(12) & -(9) & 0.88(7) & 1.07(10) & 1.13(9) \\
    Multi-stage2 & 0.53(4) & 154(12) & -(10) & 0.74(6) & 0.52(5) & 0.53(4) \\
    Log Logistic & 1.70(14) & 109(8) & -(7) & 1.47(12) & 0.19(2) & 1.70(14) \\
    Log Probit & 0.57(5) & 128(10) & -(8) & 0.52(4) & 0.03(0) & 0.57(5) \\
    Dichotomous-Hill & 0.15(1) & 42.5(3) & -(1) & 0.21(2) & 0(0) & 0.15(1) \\\hline
    \multicolumn{7}{c}{\texttt{ToxicR} informative prior} \\ \hline
    Logistic & 17.6(27) & 230(18) & -(22) & 17.4(29) & 18.3(51) & 17.6(27) \\
    Probit & 9.42(14) & 157(12) & -(15) & 9.44(16) & 10.1(28) & 9.42(14) \\
    Q-linear & 17.8(27) & 340(26) & -(30) & 17.4(29) & 4.10(11) & 17.8(27) \\
    Weibull & 8.94(14) & 152(12) & -(11) & 7.47(13) & 0.67(2) & 8.94(14) \\
    Multi-stage2 & 9.86(15) & 154(12) & -(7) & 5.40(9) & 2.62(7) & 9.84(15) \\
    Log Logistic & 0.33(1) & 109(8) & -(4) & 0.31(1) & 0.08(0) & 0.33(1) \\
    Log Probit & 1.68(3) & 128(10) & -(9) & 1.56(3) & 0.28(1) & 1.68(3) \\
    Dichotomous-Hill & 0.23(0) & 42.5(3) & -(3) & 0.20(0) & 0.78(0) & 0.23(0) \\\hline
        \multicolumn{7}{c}{Data based informative prior} \\ \hline
    Logistic & 12.3(31) & 230(18) & -(20) & 0.04(0) & 12.5(32) & 12.3(31) \\
    Probit & 7.76(19) & 157(12) & -(14) & 2.20(0) & 8.09(21) & 7.76(19) \\
    Q-linear & 5.95(15) & 340(26) & -(27) & 0.61(0) & 6.41(17) & 5.95(15) \\
    Weibull & 2.83(7) & 152(12) & -(10) & 0.55(0) & 1.25(3) & 2.83(7) \\
    Multi-stage2 & 6.54(16) & 154(12) & -(13) & $-^*$ (0) & 5.43(14) & 6.51(16) \\
    Log Logistic & 2.18(5) & 109(8) & -(7) & 3.35(0) & 2.32(6) & 2.18(5) \\
    Log Probit & 2.42(6) & 128(10) & -(8) & 3.08(0) & 2.47(6) & 2.42(6) \\
    Dichotomous-Hill & 0.07(0) & 42.5(3) & -(2) & +(100) & 0(0) & 0.06(0) \\\hline
    \end{tabular}
\end{table}

\begin{table}[H]
    \caption{Approximation of ML $\times 10^{-6}$ and weights (\%, in parentheses) of the model for Dataset 4. All results $< 10^{-9}$ are listed as 0. All results $> 10^{-3}$ are listed as ``+''. ``-'' indicates that ML is not calculated. ``$-^*$'' indicates non-computable value.}
    \label{tb:7}
    \begin{tabular}{>{\centering\arraybackslash}m{0.23\textwidth-2\tabcolsep}>{\centering\arraybackslash}m{0.13\textwidth-2\tabcolsep}>{\centering\arraybackslash}m{0.125\textwidth-2\tabcolsep}>{\centering\arraybackslash}m{0.125\textwidth-2\tabcolsep}>{\centering\arraybackslash}m{0.13\textwidth-2\tabcolsep}>{\centering\arraybackslash}m{0.13\textwidth-2\tabcolsep}>{\centering\arraybackslash}m{0.13\textwidth-2\tabcolsep}}
    \hline
     \multicolumn{7}{c}{BBMD non-informative prior} \\ \hline
      Model & Reference value & MLE-based  Schwarz criterion & MCMC-based Schwarz criterion & Laplace approximation & Density estimation & Bridge sampling \\ \hline
    Logistic & 0.56(1) & 875(20) & -(26) & 0.54(2) & 0.56(3) & 0.56(1) \\
    Probit & 0.18(0) & 910(21) & -(27) & 0.18(1) & 0.18(1) & 0.18(0) \\
    Q-linear & 0.63(1) & 449(10) & -(14) & 0.62(3) & 0.64(3) & 0.63(1) \\
    Weibull & 3.28(7) & 474(11) & -(4) & $-^*$ (0) & 3.49(17) & 3.29(7) \\
    Multi-stage2 & 0(0) & 469(11) & -(13) & 0.02(0) & 0.01(0) & 0(0) \\
    Log Logistic & 7.12(15) & 481(11) & -(6) & 2.74(12) & 6.47(32) & 7.10(15) \\
    Log Probit & 3.65(8) & 481(11) & -(5) & 0.97(4) & 2.34(12) & 3.68(8) \\
    Dichotomous-Hill & 31.8(67) & 237(5) & -(4) & 17.0(77) & 6.65(33) & 31.6(67) \\\hline
    \multicolumn{7}{c}{\texttt{ToxicR} informative prior} \\ \hline
    Logistic & 6.96(2) & 875(20) & -(15) & 6.80(2) & 7.06(15) & 6.98(2) \\
    Probit & 24.7(6) & 910(21) & -(22) & 24.4(7) & 26.4(56) & 24.8(7) \\
    Q-linear & 49.0(13) & 449(10) & -(14) & 47.6(14) & 2.44(5) & 49.1(13) \\
    Weibull & 62.5(16) & 474(11) & -(11) & 56.2(16) & 1.38(3) & 62.4(16) \\
    Multi-stage2 & 14.3(4) & 469(11) & -(10) & 14.6(4) & 0.56(1) & 14.3(4) \\
    Log Logistic & 124(33) & 481(11) & -(12) & 115(33) & 4.99(11) & 124(32) \\
    Log Probit & 71.6(19) & 481(11) & -(10) & 61.2(18) & 4.48(9) & 71.5(19) \\
    Dichotomous-Hill & 28.3(7) & 237(5) & -(5) & 23.5(7) & 0(0) & 28.3(7) \\\hline
        \multicolumn{7}{c}{Data based informative prior} \\ \hline
    Logistic & 215(15) & 875(20) & -(22) & 97.9(12) & 227(21) & 215(15) \\
    Probit & 235(16) & 910(21) & -(23) & 248(31) & 238(22) & 235(16) \\
    Q-linear & 119(8) & 449(10) & -(11) & 70.3(9) & 113(10) & 119(8) \\
    Weibull & 145(10) & 474(11) & -(9) & 113(14) & 86.5(8) & 145(10) \\
    Multi-stage2 & 191(13) & 469(11) & -(10) & $-^*$ (0) & 237(22) & 193(14) \\
    Log Logistic & 146(10) & 481(11) & -(9) & 144(18) & 83.7(8) & 146(10) \\
    Log Probit & 148(10) & 481(11) & -(9) & 131(16) & 98.4(9) & 148(10) \\
    Dichotomous-Hill & 232(16) & 237(5) & -(5) & $-^*$ (0) & 3.61(0) & 229(16) \\\hline
    \end{tabular}
\end{table}

\subsection{Estimation of BMD and calculation of BMDL}

\begin{table}[H]
    \caption{BMD estimates and BMDL values (in parentheses). Upper table: Results on applying BBMD non-informative prior; lower table: results on applying \texttt{ToxicR}-informative prior.}
    \label{tb:8}
    \begin{tabular}{>{\centering\arraybackslash}m{0.1\textwidth-2\tabcolsep}>{\centering\arraybackslash}m{0.13\textwidth-2\tabcolsep}>{\centering\arraybackslash}m{0.13\textwidth-2\tabcolsep}>{\centering\arraybackslash}m{0.13\textwidth-2\tabcolsep}>{\centering\arraybackslash}m{0.13\textwidth-2\tabcolsep}>{\centering\arraybackslash}m{0.13\textwidth-2\tabcolsep}>{\centering\arraybackslash}m{0.13\textwidth-2\tabcolsep}>{\centering\arraybackslash}m{0.12\textwidth-2\tabcolsep}}
     \multicolumn{8}{c}{BBMD non-informative prior} \\ \hline
     Dataset & Reference value & MLE-based Schwarz criterion & MCMC-based Schwarz  criterion & Laplace approximation & Density estimation & Bridge sampling & Equal weight \\ \hline
     1 & 0.270\par(0.194) & 0.243\par(0.174) & 0.229\par(0.167) & 0.201\par(0.150) & 0.268\par(0.193) & 0.271\par(0.194) & 0.209\par(0.151) \\
    2 & 0.456\par(0.308) & 0.573\par(0.319) & 0.486\par(0.330) & 0.567\par(0.463) & 0.601\par(0.329) & 0.456\par(0.308) & 0.568\par(0.330) \\
    3 & 0.033\par(0.015) & 0.040\par(0.019) & 0.037\par(0.020) & 0.032\par(0.016) & 0.025\par(0.015) & 0.033\par(0.015) & 0.049\par(0.019) \\
    4 & 0.541\par(0.258) & 0.476\par(0.247) & 0.387\par(0.233) & 0.487\par(0.249) & 0.624\par(0.267) & 0.542\par(0.258) & 0.499\par(0.241) \\ \hline
    \multicolumn{8}{c}{\texttt{ToxicR} informative prior} \\ \hline
    1 & 0.173\par(0.120) & 0.169\par(0.119) & 0.175\par(0.124) & 0.172\par(0.120) & 0.165\par(0.115) & 0.173\par(0.120) & 0.151\par(0.107) \\
    2 & 0.423\par(0.273) & 0.409\par(0.283) & 0.413\par(0.291) & 0.421\par(0.272) & 0.508\par(0.404) & 0.423\par(0.273) & 0.422\par(0.293) \\
    3 & 0.036\par(0.022) & 0.037\par(0.020) & 0.036\par(0.021) & 0.035\par(0.023) & 0.039\par(0.028) & 0.036\par(0.022) & 0.039\par(0.018) \\
    4 & 0.345\par(0.190) & 0.324\par(0.205) & 0.319\par(0.200) & 0.342\par(0.189) & 0.342\par(0.223) & 0.345\par(0.190) & 0.321\par(0.191) \\ \hline
    \multicolumn{8}{c}{Data based informative prior} \\ \hline
    1 & 0.176\par(0.127) & 0.177\par(0.127) & 0.179\par(0.129) & 0.167\par(0.118) & 0.175\par(0.126) & 0.176\par(0.127) & 0.156\par(0.108) \\
    2 & 0.391\par(0.261) & 0.398\par(0.273) & 0.398\par(0.277) & 0.531\par(0.431) & 0.388\par(0.268) & 0.391\par(0.261) & 0.411\par(0.286) \\
    3 & 0.034\par(0.022) & 0.031\par(0.018) & 0.032\par(0.019) & 0.037\par(0.007) & 0.035\par(0.023) & 0.034\par(0.022) & 0.033\par(0.015) \\
    4 & 0.330\par(0.192) & 0.324\par(0.197) & 0.323\par(0.201) & 0.325\par(0.196) & 0.319\par(0.197) & 0.330\par(0.192) & 0.321\par(0.182) \\ \hline
    \end{tabular}
\end{table}

\begin{figure}[H]
    \centering
    \includegraphics[width=0.85\textwidth]{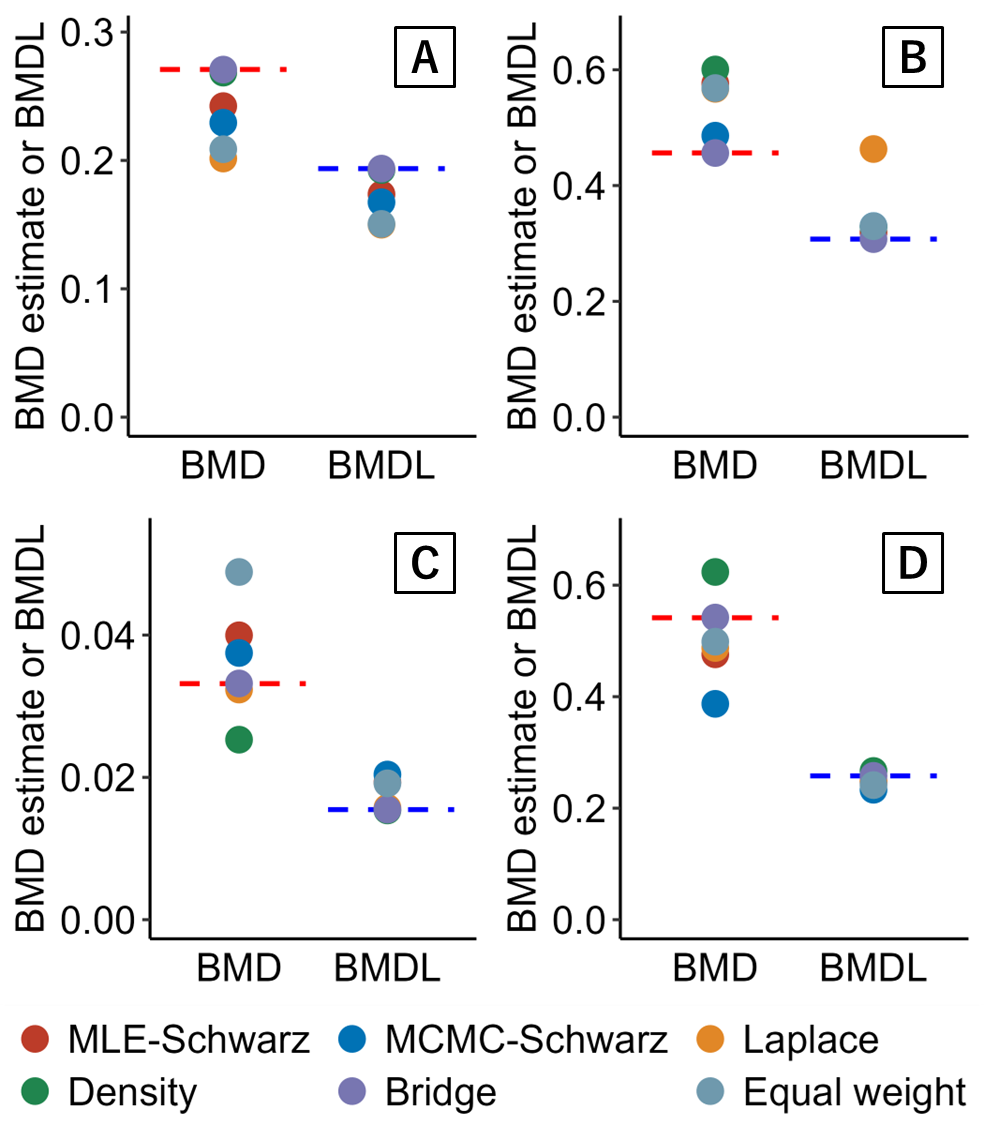}
    \caption{BMD estimates and BMDLs using BBMD non informative prior. Panel A, Dataset 1; Panel B, Dataset 2; Panel C, Dataset 3; and Panel D, Dataset 4. The red-dashed line is the BMD reference value, and the blue-dashed line is the BMDL reference value. MLE: Schwarz criterion using the maximum likelihood; MCMC: Markov chain Monte Carlo.}
    \label{fig:1}
\end{figure}

\begin{figure}[H]
    \centering
    \includegraphics[width=0.85\textwidth]{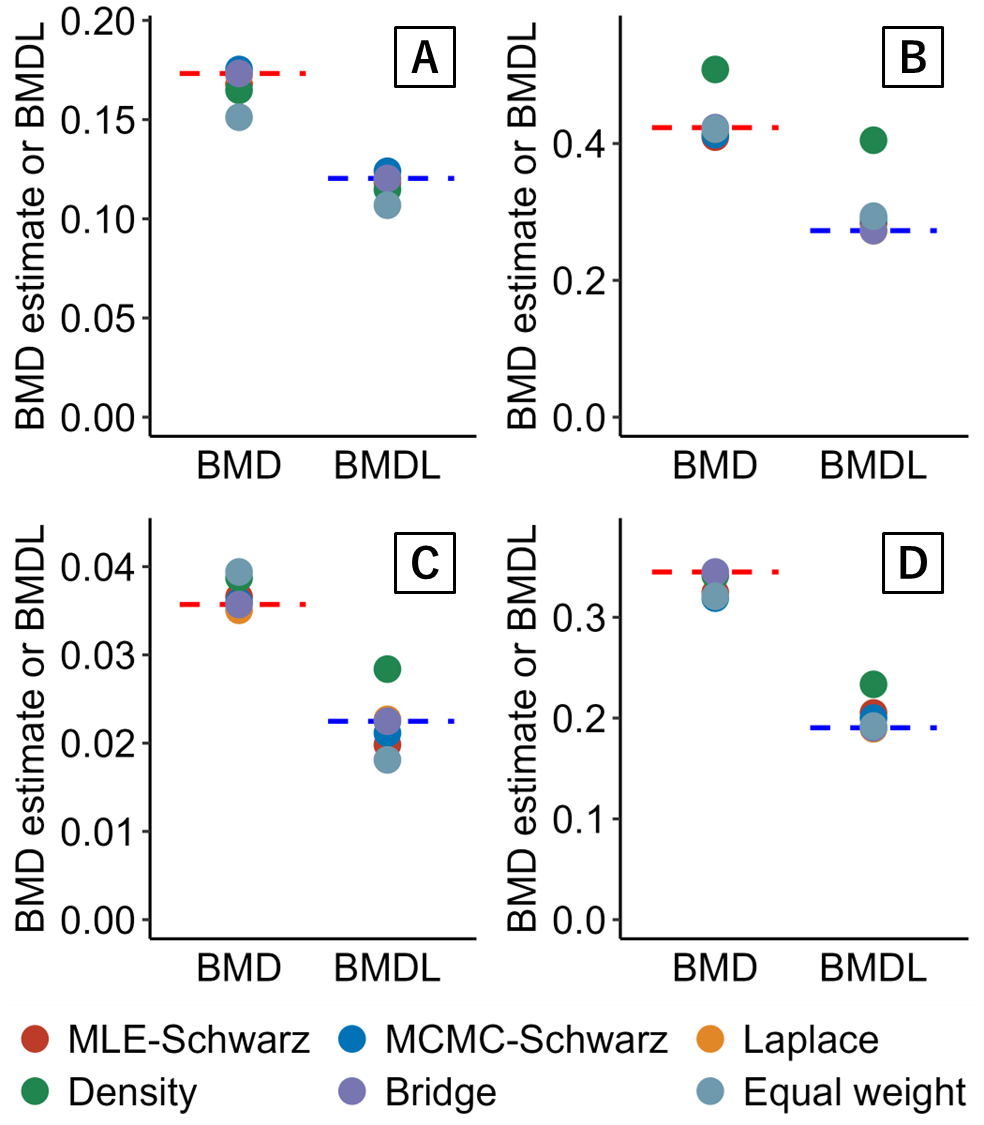}
    \caption{BMD estimates and BMDLs using \texttt{ToxicR} informative prior. Panel A, Dataset 1; Panel B, Dataset 2; Panel C, Dataset 3; and Panel D, Dataset 4. The red-dashed line is the BMD reference value, and the blue-dashed line is the BMDL reference value. MLE: Schwarz criterion using the maximum likelihood; MCMC: Markov chain Monte Carlo.}
    \label{fig:2}
\end{figure}

\begin{figure}[H]
    \centering
    \includegraphics[width=0.85\textwidth]{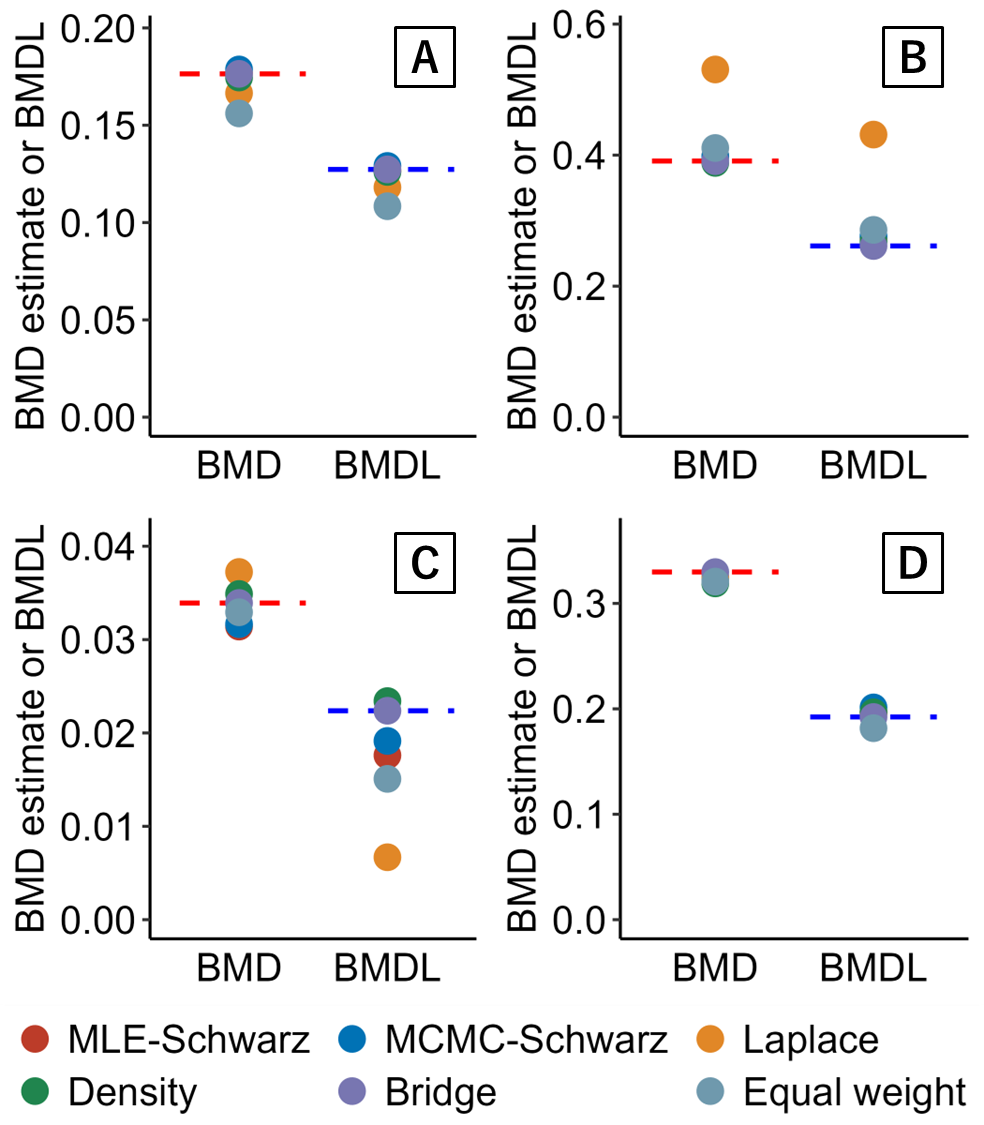}
    \caption{BMD estimates and BMDLs using \texttt{ToxicR} informative prior. Panel A, Dataset 1; Panel B, Dataset 2; Panel C, Dataset 3; and Panel D, Dataset 4. The red-dashed line is the BMD reference value, and the blue-dashed line is the BMDL reference value. MLE: Schwarz criterion using the maximum likelihood; MCMC: Markov chain Monte Carlo.}
    \label{fig:3}
\end{figure}